\documentclass[a4paper,reqno,12pt]{amsart}
\usepackage{amsfonts,amssymb,amsthm,bbm,amscd}
\usepackage{graphicx,epsfig}
\usepackage{perpage}
\usepackage{url}
\usepackage{color}
\usepackage{epstopdf}
\usepackage{mathabx}
\usepackage{amsmath}
\usepackage{breqn}
\usepackage{amssymb}
\usepackage{amsfonts}
\usepackage{float}
\usepackage{framed}
\usepackage{subfigure}
\usepackage{mathtools}
\usepackage{enumerate}
\usepackage{hyperref}
\usepackage{microtype}
\usepackage{comment}
\usepackage{mathabx}

\usepackage{palatino}
\usepackage[T1]{fontenc}
\usepackage[mathscr]{eucal}
\usepackage{dsfont}

\usepackage{fullpage}

\usepackage[font=small,labelfont=bf]{caption}
\usepackage{subfigure}
\usepackage{tikz}
\usetikzlibrary{calc,patterns}

\usepackage{hyperref}
\hypersetup{
colorlinks=true,
linktocpage=true,
pdfstartview=FitH,
breaklinks=true,
pdfpagemode=UseNone,
pageanchor=true,
pdfpagemode=UseOutlines,
plainpages=false,
bookmarksnumbered,
bookmarksopen=false,
bookmarksopenlevel=1,
hypertexnames=true,
pdfhighlight=/O,
pdftitle={},
pdfauthor={},
pdfsubject={},
pdfkeywords={},
pdfcreator={pdfLaTeX},
pdfproducer={LaTeX with hyperref}
}

\numberwithin{equation}{section}  
\usepackage[sort&compress,capitalize,nameinlink]{cleveref}
\crefname{app}{Appendix}{Appendices}

\crefrangeformat{equation}{\upshape(#3#1#4)\textendash(#5#2#6)}

\theoremstyle{plain}
\newtheorem{theorem}{Theorem}

\numberwithin{theorem}{section}

\allowdisplaybreaks

\begin{document}

\title{Exploring Metastability in Ising models: \\ critical droplets, energy barriers and exit time.}

\date{}

\author[V. Jacquier]{Vanessa Jacquier}

\begin{abstract}
This paper provides an overview of the research on the metastable behavior of the Ising model. We analyze the transition times from the set of metastable states to the set of the stable states by identifying the critical configurations that the system crosses with high probability during this transition and by computing the energy barrier that the system must overcome to reach the stable state starting from the metastable one. We describe the dynamical phase transition of the Ising model evolving under Glauber dynamics across various contexts, including different lattices, dimensions and anisotropic variants. The analysis is extended to related models, such as long-range Ising model, Blume-Capel and Potts models, as well as to dynamics like Kawasaki dynamics, providing insights into metastability across different systems.
\end{abstract}

\maketitle


\section{Introduction}
The Ising model is one of the most interesting models in statistical physics, introduced by Ernst Ising in 1925 to describe the behavior of magnetic systems. The model consists on a network of interacting particles or spins that take value $+1$ or $-1$. We can imagine each vertex of the lattice as an atom with its magnetic moment (or spin) capable of aligning in one of two directions: up or down. The interaction between neighboring spins is described by a Hamiltonian of the same form given in \eqref{hamiltonianFunction}. In particular, the energy of the system depends, not only on the number of pluses and minuses, but also on the interaction between neighboring vertices that give different energy contributions depending on whether the spins are the same or different.

One of the most captivating features of the Ising model is the phase transition that occurs when the temperature of the system changes. At high temperatures, the spins tend to orient randomly, and the system is in a disordered (\emph{paramagnetic}) state. At lower temperatures, the spins tend to align, and the system enters an ordered (\emph{ferromagnetic}) state. The temperature at which this transition occurs is called the critical temperature $T_c$. In a regime near $T_c$, the Ising model exhibits a phenomenon known as metastability. 
Metastability is an ubiquitous phenomenon in nature, which appears in a plethora of diverse fields including physics, chemistry, biology, computer science, climatology and economics.

More formally, metastability is described as a dynamical phenomenon that occurs when a system is close to a first order phase transition. After changing some thermodynamic parameters, the system remains for a considerable (random) time in its previous phase, the metastable state, before suddenly making a transition to the new phase, the stable state. 
In other words, on a short time scale, the system behaves as if it was in equilibrium, while, on a long time scale, it moves between different regions of the state space. At low temperature, this motion is preceded by the appearance of a \emph{critical} mesoscopic configuration of the system via a spontaneous fluctuation or some external perturbation. 
Thus, when the system is initiated in the metastable phase, it starts its long transition towards the stable phase. In particular, it must overcome an energy barrier to reach the stable state starting from the metastable state. The detailed evolution of the system is based an Hamiltonian function and the associated dynamics. Moreover, it is possible to define an equilibrium measure in terms of the Hamiltonian, for example the Gibbs measure. 

Metastability typically raises three key questions. 
The first involves studying the \emph{transition time} from the set of metastable states to the set of the stable states, i.e., the time needed to arrive at the equilibrium phase. Although the fluctuations of the dynamics trigger the transition, these are very unlikely, so the system is typically stuck in the metastable state for an exponentially long time. 
This long wait is also due to multiple failed attempts to escape the metastable state.
The second question concerns the identification of the so-called \emph{critical configurations} that the system creates in order to reach equilibrium. Roughly speaking, the system fluctuates around the metastable state until it visits a critical configuration which then allows it to finally reaches equilibrium. 
The third question focuses on the study of the typical paths that the system follows with high probability during the transition from the metastable state to the stable state. These are the so-called \emph{tube of typical trajectories}.

To understand how the transition from a mestastable state to the stable state takes place, it is necessary to explore the specific features of the model under consideration. For instance, the shape of the critical configurations depends on the temperature and the geometrical structure of the configurations (e.g., anisotropy effects or the features of the type of dynamics in the case of discrete models). 

There are different rigorous dynamical approaches to study metastability. The first approach, known as \emph{pathwise approach}, was first initiated in \cite{cassandro1984metastable} and further developed in \cite{ scoppola1994metastability, olivieri1995markov, olivieri1996markov}, see also \cite{olivieri2005large} as a general reference. 
This approach provides large deviation estimates for the first hitting time of the stable set and the tube of typical trajectories.  It has been successfully applied to the rigorous analysis of metastability in various contexts, such as in the infinite volume limit, at low temperatures, or with vanishing magnetic fields. 
Central to this method are the notions of cycles and cycle paths and it hinges on a detailed knowledge of the energy landscape. Similar results based on a graphical representation of cycles were independently obtained in \cite{catoni1997exit, catoni1999simulated}, and subsequently applied to reversible Metropolis dynamics and simulated annealing in \cite{catoni1992parallel, trouve1996rough}. 
The pathwise approach was further developed in \cite{manzo2004essential, cirillo2013relaxation, cirillo2015metastability, fernandez2015asymptotically, fernandez2016conditioned} to disentangle the study of transition time from the one of typical trajectories and to treat irreversible dynamics. This framework offers a clear physical interpretation of the metastable state and of the associated exit time. Indeed, in the low temperature limit, the time for the system to leave a neighborhood of the metastable state is exponentially long and the exponential rate is proportional to the inverse temperature and to the minimal energy barrier that the system must overcome to reach the stable state.
Additionally, this approach yields insights into the likely paths the system follows during its transition to the stable state. Notably, before reaching the stable state, with high probability the system visits one of the critical configurations where the smallest energy barrier is attained.

Another successful approach is the \emph{potential-theoretic approach}, initiated in \cite{bovier2002metastability}. This theory provides precise estimates of the expected value of the transition time from the metastable to the stable states. For a comprehensive overview and applications to various models, see \cite{bovier2016metastability}.
Crucially, this method allows the computation of the prefactor of the mean transition time. Indeed, the transition time can be written in terms of a quantity known as the \emph{capacity}, which can be bounded from above and from below by exploiting appropriate variational principles. However, applying these principles in practice can be demanding, as it often requires detailed information about the critical configurations involved, as discussed in \cite{bovier2004metastability, bovier2016metastability}. This theory reveals that the prefactor depends on the structure of the critical droplets that the system crosses during the transition to the stable state. 
The potential-theoretic approach has been successfully used to analyze metastable behavior under Metropolis dynamics in \cite{boviermanzo2002metastability, bovier2006sharp, den2012metastability, cirillo2017sum, bashiri2017note}, among others.

Other approaches have been proposed in \cite{beltran2010tunneling, beltran2012tunneling, gaudillierelandim2014, landim2023resolvent} and in \cite{bianchi2016metastable}. 
In \cite{beltran2010tunneling}, the authors introduce definitions of tunneling and metastability for continuous-time Markov processes on countable state spaces.  They derive sufficient conditions under which an irreducible, positive recurrent Markov process exhibits tunneling behavior. For reversible dynamics, these conditions can be characterized using potential theory the martingale approach (see \cite{landim2019metastability} for details), and can be expressed in terms of the capacities and of the stationary measure of the process. Subsequently, in \cite{beltran2012tunneling}, this framework is extended to non-reversible dynamics, leveraging the Dirichlet principle established in \cite{gaudillierelandim2014}.

A more recent method, the so-called \emph{resolvent approach}, was developed in \cite{landim2023resolvent}. In this work, the authors show that the metastable behavior of a sequence of Markov chains can be determined by examining a specific property of the solutions to the resolvent equation associated with the process generator. Remarkably, this property is not only sufficient to characterize metastability but also necessary. Since these conditions for metastability do not require explicit knowledge of the stationary state, they can, in principle, be applied to non-reversible dynamics where the stationary state is unknown. It is worth noting that these necessary and sufficient conditions for metastability can be derived from those presented in \cite{beltran2010tunneling, beltran2012tunneling}.

Finally, in \cite{bianchi2016metastable} the authors investigate metastability for Markov chains on finite configuration spaces within certain asymptotic regimes, building on the approaches of \cite{penrose1971rigorous, penrose1987towards}. By comparing restricted ensembles and quasi-stationary measures, and introducing \emph{soft measures} as a conceptual bridge between the two, they prove an asymptotic exponential exit law and, on a generally different time scale, an asymptotic exponential transition law. Leveraging tools from potential theory and defining a specialized form of capacities, they provide precise estimates for relaxation times, mean exit times, and transition times.

In this paper, we summarize the main results concerning the metastable behavior of the Ising model. Section \ref{sec:model1} provides a detailed description of the Ising model and the Glauber dynamics governing its evolution. In particular, we outline the model-independent definitions and key results related to metastability phenomenon (Sections \ref{metastability_main_tools} and \ref{metastability_main_results}). Additionally, we reformulate the Hamiltonian function in terms of the perimeter and area of specific geometrical figures, the \emph{polyominoes}, as detailed in Section \ref{sec:polyominos}.
Subsequently, we present the main findings regarding the metastable behavior of the standard 2D Ising model on various structures, including the square lattice (Section \ref{subsec:square}), the hexagonal lattice (Section \ref{subsec:hexagonal}), and (random) graphs (Section \ref{subsec:random_graph}).
In Section \ref{sec:3D}, we discuss the metastable behavior of the Ising model in higher dimensions. Section \ref{sec:aniso} explores the primary outcomes for anisotropic variants of the Ising model, while Section \ref{subsec:longrange} introduces the long-range Ising model, characterized by interactions between all spins.
Finally, Section \ref{sec:extension} examines metastability in several extensions of the Ising model, including the Blume-Capel model (Section \ref{blume_capel}) and the Potts model (Section \ref{sec:Potts}). We also report the main results for the evolution of the Ising model under conservative dynamics, specifically Kawasaki dynamics, in Section \ref{sec:Kawasaki}.

\section{Ising model with Glauber dynamics}
\label{sec:model1}
Consider a subset $\Lambda$ of a discrete lattice 
and associate a spin variable $\sigma(i)\in \{-1,+1\}$ to each site $i\in\Lambda$. We interpret $\sigma(i)=+1$ (respectively $\sigma(i)=-1$) as indicating that the spin at site $i$ is pointing upwards (respectively downwards). 

On the \emph{configuration space} $\mathcal{X}:=\{-1,+1\}^{\Lambda}$ we consider the \emph{Hamiltonian function} $H: \mathcal{X} \longrightarrow \mathbb{R}$ defined as
\begin{equation}\label{hamiltonianFunction}
H(\sigma):=-\frac{J}{2}\sum_{\substack{i,j \in \Lambda\\ d(i, j)=1}} \sigma (i) \sigma (j) -\frac{h}{2} \sum_{i \in \Lambda} \sigma (i),
\end{equation} 
where $J>0$ represents the ferromagnetic interaction between two spins, $h>0$ is the external magnetic field and $d(\cdot, \cdot)$ is the lattice distance. 
The parameter $h$ is chosen \emph{small enough} to ensure that the system exhibits a metastable behavior.
In the following we denote by $\textbf{+1}$ (resp. $\textbf{-1}$) the configuration $\sigma$ such that $\sigma(i)=+1$ (resp. $\sigma(i)=-1$) for every $i\in\Lambda$.

In order to study the evolution of the system, we introduce \emph{Glauber dynamics}. We consider a Markov chain  $(\sigma_t)_{t \in \mathbb{N}}$ on $\mathcal{X}$ defined via the so called \emph{Metropolis Algorithm} where the
transition probabilities between two configurations $\sigma$ and $\eta$ are given by
\begin{equation}\label{def:glauber}
    p(\sigma, \eta) \footnote{$[x]_+$ denotes the positive part of $x$, i.e., $[x]_+=x$ if $x>0$ and $[x]_+=0$ otherwise.}
    =\left\{
    \begin{array}{ll}
    q(\sigma,\eta) e^{-\beta[H(\eta) -H(\sigma)]_+} &\qquad \text{if } \sigma \neq \eta \\
    1- \sum_{\eta \in \mathcal{X}} q(\sigma,\eta) e^{-\beta[H(\eta) -H(\sigma)]_+} &\qquad \text{if } \sigma =\eta
    \end{array}
    \right.
\end{equation}
where the parameter $\beta:=\frac{1}{T} >0$ is the inverse temperature,
and $q(\sigma,\eta)$ is a connectivity matrix independent of $\beta$, defined as
\begin{equation}
    q(\sigma,\eta)= \left\{
    \begin{array}{ll}
    \frac{1}{|\Lambda|} & \qquad \textrm{ if } \exists \, x\in \Lambda: \sigma^{(x)}=\eta, \\
    0 & \qquad \textrm{ otherwise, }
    \end{array}
    \right.
\end{equation}
with
\begin{equation}\label{sigma_x}
    \sigma^{(x)}(z)= \left\{
    \begin{array}{ll}
        \sigma(z) & \;\;\textrm{ if } z \neq x,\\
        -\sigma(x) & \;\;\textrm{ if } z = x.
        \end{array}
    \right.
\end{equation}

It is possible to check that $(\sigma_t)_{t \in \mathbb{N}}$  is an ergodic-aperiodic Markov chain  on $\mathcal{X}$  satisfying the detailed balance condition
\begin{equation}\label{reversibility}
    \mu(\sigma)p(\sigma, \eta)=\mu(\eta)p(\eta, \sigma),
\end{equation}
with respect to the Gibbs measure 
\begin{equation}\label{def:gibbs}
    \mu(\sigma)=\frac{e^{-\beta H(\sigma)}}{\sum_{\eta \in \mathcal{X}} e^{-\beta H(\eta)}}. 
\end{equation}
This implies that the measure $\mu$ is stationary, that is $\sum_{\sigma\in \mathcal{X}}\mu(\sigma)p(\sigma,\eta)=\mu(\eta)$.

\subsection{Metastability: main tools}\label{metastability_main_tools} The problem of metastability is the study of the first arrival of the process $(\sigma_t)_{t \in \mathbb{N}}$ to the set of the \emph{stable states}, corresponding to the set of absolute minima of the Hamiltonian function, starting from an initial local minimum. 
The local minima can be distinguished by their stability level, i.e., the height of the energy barrier separating them from lower energy states. 
More precisely, for any $\sigma \in \mathcal{X}$, let $\mathcal{I}_{\sigma}$ be the set of configurations with energy strictly lower than $H(\sigma)$, i.e.,
\begin{equation}\label{I}
\mathcal{I}_{\sigma}:=\{\eta\in \mathcal{X} \,|\, H(\eta)<H(\sigma)\}.
\end{equation}
Let $\omega=\{\omega_1,\ldots,\omega_m\}$ be a finite sequence of configurations in $\mathcal{X}$ such that $\omega_{k+1}$ is obtained from $\omega_k$ by a single spin flip, for each $k$ from $1$ to $m-1$. We call $\omega$ a \emph{path} with starting configuration $\omega_1$ and final configuration $\omega_m$ and we denote it by $\omega\colon\, \omega_1 \to \omega_m$. The configurations composing $\omega$ are called \emph{connected configurations}. Moreover, we indicate the set of all these paths as $\Theta(\omega_1,\omega_m)$. We note that this definition of the path is different if we choose different dynamics.

\begin{figure}[!htb]
        \begin{center}
        \includegraphics[scale=1.3]{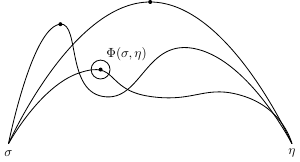} \,\,\, \includegraphics[scale=1.3]{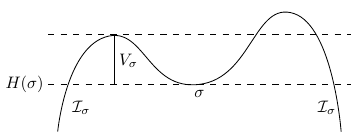}
        \end{center}
        \caption{On the left, an example of the communication height between two configurations $\sigma$ and $\eta$. On the right, the stability level of a configuration $\sigma$.}
        \label{fig:comm_height}
    \end{figure}

We call \emph{communication height} between two configurations $\sigma$ and $\eta$ the minimum among all maximal energies along the paths in $\Theta(\sigma,\eta)$, i.e.,
\begin{equation}\label{minmax}
\Phi(\sigma,\eta):=\min_{\omega\in\Theta(\sigma,\eta)}\max_{\zeta \in \omega} H(\zeta).
\end{equation}
See the left panel in Figure \ref{fig:comm_height}. 

By $\Phi(\omega)$ we denote the communication height along the path 
\mbox{$\omega=\{\omega_1,\ldots,\omega_n\}$}, i.e. \linebreak 
\mbox{$\Phi(\omega)=\max_{i=1,\ldots,n} H(\omega_i)$}. 
Similarly, we also define the \emph{communication height} 
between two sets $A, B \subset \mathcal{X}$ as
\begin{equation}
\Phi(A,B):=\min_{\sigma \in A,\eta \in B} \Phi(\sigma,\eta).
\end{equation}

Let $\mathcal{X}^s$ be the set of global minima of the energy, and we refer to it as the set of the stable states (or ground states).
The \emph{stability level} of a configuration $\sigma \not \in \mathcal{X}^s$ is defined as follows
\begin{equation}
V_{\sigma}:=\Phi(\sigma,\mathcal{I}_\sigma)-H(\sigma).
\end{equation}
If there are no configurations with energy smaller than $H(\sigma)$, then we set $V_{\sigma}=\infty$. 
We note that the stability level $V_\sigma$ is the minimal cost that, starting from $\sigma$, has to be payed in order to reach states at energy lower than $H(\sigma)$.
See the right panel in Figure \ref{fig:comm_height}. 

The stability level plays a crucial role to describe the critical configurations and to identify the metastable states. Thus, it is useful to characterize the configurations in terms of their stability level, in particular given a positive value $V$ we define the set of all configurations with stability level strictly greater than $V$, i.e. 

\begin{equation}\label{Xv}
\mathcal{X}_V:=\{x\in \mathcal{X} \,\, | \,\, V_{x}>V\}.
\end{equation}

To define the set of metastable states, we introduce the \emph{maximal stability level} 
\begin{equation}\label{Gamma}
    \Gamma_m:=\max_{\sigma\in \mathcal{X}\setminus \mathcal{X}^s}V_{\sigma}.
\end{equation}

The metastable states are those state that attain the maximal stability level $\Gamma_m< \infty$, i.e. \begin{align}
    \mathcal{X}^m:=\{\sigma \in \mathcal{X}| \, V_{\sigma}=\Gamma_m \}.
\end{align}
In \cite{cirillo2013relaxation}, the authors find a relaxation between the maximal stability level $\Gamma_m$ and the communication height between a metastable state \textbf{m} and a stable state \textbf{s}, i.e.
\begin{align}
    \max_{\sigma\in \mathcal{X}\setminus \mathcal{X}^s}V_{\sigma}=\Phi (\textbf{m}, \textbf{s})-H(\textbf{m}).
\end{align}
Thus, in the rest of the paper, we will use the notation $\Gamma$ to refer indistinctly to the two definitions and we call it the \emph{energy barrier} of the system.

We frame the problem of metastability as the identification of metastable states 
and the computation of transition times from the metastable states to the stable ones. 
To study the transition between $\mathcal{X}^m$ and $\mathcal{X}^s$, 
we define the \emph{first hitting time} of 
$A\subset \mathcal{X}$ starting from $\sigma \in \mathcal{X}$
\begin{equation}\label{fht}
    \tau^{\sigma}_A:=\inf\{t>0 \,|\, X_t\in A\}.
\end{equation}
Whenever possible we shall drop the superscript denoting the
starting point $\sigma$ from the notation and we denote by $\mathbb{P}_{\sigma}(\cdot)$ and $\mathbb{E}_{\sigma}[\cdot]$ respectively the probability and the average along the trajectories of the process started at $\sigma$. 

In order to describe the evolution of the system and the shape of the critical configurations, we consider the set of paths, the so-called \emph{optimal paths}, realizing the minimal value of the maximal energy in the paths between any metastable state and the set of the stable states. Formally, we define the set of optimal paths from $\mathcal{A}$ to $\mathcal{B}$, i.e., the set of all paths from $\mathcal{A}$ to $\mathcal{B}$ realizing the min-max \eqref{minmax} in $\mathcal{X}$ between $\mathcal{A}$ and $\mathcal{B}$, and we denote it by $(\mathcal{A} \to \mathcal{B})_{opt}$.
  \begin{figure}[!hbt]
   \centering
    \includegraphics[scale=0.9]{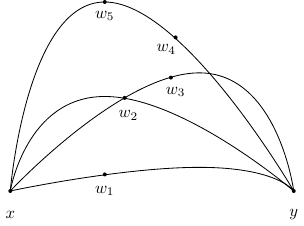}
    \caption{In the figure, we can see the optimal path $(x \to y)_{opt}$ and the set of saddles $\mathcal{S}(x,y)=\{\omega_1,...,\omega_5\}$. In this case, the minimal gates are $\{\omega_1,\omega_2,\omega_4\}$ and $\{\omega_1,\omega_2,\omega_5\}$.} \label{fig:gates_saddles}
\end{figure}
The set of the \emph{minimal saddles} between two configurations $\sigma, \eta \in \mathcal{X}$ is defined as the set of all maxima in the optimal paths between two configurations, i.e.
\begin{equation}
    \mathscr{S}(\sigma,\eta):=\{\zeta \in \mathcal{X} \, | \, \exists \, \omega:\sigma \to \eta, \, \omega \ni \zeta \text{ such that } \max_{\xi \in \omega} H(\xi)=H(\zeta)=\Phi(\sigma,\eta)\},
\end{equation}
and define
\begin{equation}
\mathscr{S}(\mathcal{A},\mathcal{B}):= \bigcup_{\substack{\sigma \in \mathcal{A}, \, \eta \in \mathcal{B}: \\ \Phi(\sigma,\eta)=\Phi(\mathcal{A},\mathcal{B})}} \mathscr{S}(\sigma,\eta).
\end{equation}
We focus on the subsets of saddles that are typically visited during the last excursion from a metastable state to the set of the stable states. To this end, we introduce the \emph{gates} from metastability to stability, defined as the subsets of $\mathscr{S}$ visited by all the optimal paths. More precisely,
given a pair of configurations $\sigma, \eta \in \mathcal{X}$, we say that $\mathcal{W} \equiv \mathcal{W}(\sigma, \eta)$ is a gate for the transition from $\sigma$ to $\eta$ if $\mathcal{W}(\sigma, \eta) \subseteq \mathscr{S}(\sigma,\eta)$ and $\omega \cap \mathcal{W} \neq \emptyset$ for all $\omega \in (\sigma \to \eta)_{opt}$.  
Moreover, we introduce the \emph{minimal gate} that is a minimal (by inclusion) subset of $\mathscr{S}(\sigma,\eta)$ visited by all optimal paths.
The configurations in the \emph{minimal gates} have the physical meaning of \emph{critical configurations} and are central objects both from a probabilistic and from a physical point of view, since the system crosses the critical configurations in order to reach the equilibrium. Formally,
a gate $\mathcal{W}$ is a minimal gate for the transition from $\sigma$ to $\eta$ if for any $\mathcal{W}' \subset \mathcal{W}$ there exists $\omega' \in (\sigma \to \eta)_{opt}$ such that $\omega' \cap \mathcal{W}' = \emptyset$. 
For a given pair $\eta, \eta'$, there may be several disjoint minimal gates. We denote by $\mathscr{G}(\eta, \eta')$ the union of all minimal gates:
\begin{equation}
    \mathscr{G}(\eta,\eta'):= \bigcup_{\mathcal{W}: \, \text{minimal gate for } (\eta, \eta')} \mathcal{W}
\end{equation}
Obviously, $\mathscr{G}(\sigma,\sigma') \subseteq \mathscr{S}(\sigma,\sigma')$ and $\mathscr{S}(\sigma,\sigma')$ is a gate (but in general it is not minimal). The configurations $\xi \in \mathscr{S}(\eta,\eta') \setminus \mathscr{G}(\eta,\eta')$ (if any) are called \emph{dead ends}.

 Next, we classify any saddle as either unessential or essential. A saddle $z\in\mathcal{S}(x,y)$ is called \textbf{unessential} if for any $\omega\in (x\to y)_{opt}$ such that $\omega\cap z\not=\emptyset$ we have $\{ \arg \max_{\omega} H \} \setminus \{ z \}\neq\emptyset$ 
 \footnote{
 Given a function $f:\mathcal{X}\to \mathbb{R}$ and a subset 
   $A \subseteq \mathcal{X}$, we denote by 
   \begin{equation}\label{eq:arg_max_Metropolis} 
    {\arg \max}_{ A} f :=\Big\{ x\in A \, | \, f(x)=\max_{y \in A} f(y) \Big\}
    \end{equation}
   the set of points where the maximum of $f$ in $A$ is reached.
 } 
 and there exists $\omega'\in (x\to y)_{opt}$ such that 
$\{ \arg \max_{\omega'} H\}  \subseteq \{ \arg \max_{\omega} H \} \setminus \{ z \}$.

\begin{figure}[!hbt]
   \centering
    \includegraphics[scale=1]{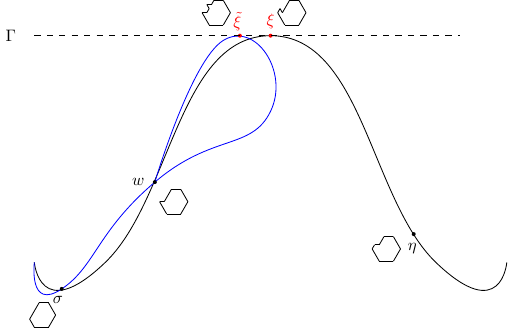}
    \caption{The configuration $\tilde \xi$ is an unessential saddle, while $\xi$ is an essential saddle. Indeed, if the system follows the blue path, starting from $\sigma$ and before reaching $\eta$, it returns to $w$ and crosses $\xi$ following the black path. This figure is taken on \cite{jacquier2022metastability}.} \label{fig:saddles_essential}
\end{figure}

A saddle  $z\in \mathcal{S} (x,y)$ is called \textbf{essential} 
if it is not unessential, i.e., if either
\begin{itemize}
\item[(i)] there exists $\omega\in(x\to y)_{opt}$ such  that 
    $\{\arg \max_\omega H\} = \{ z \}$, 
or 
\item[(ii)] there exists $\omega\in(x\to y)_{opt}$ such that 
     $\{\arg \max_\omega H\} \supset \{ z \}$ 
     and \\
     $\{\arg \max_{\omega'} H\} 
      \not \subseteq \{ \arg \max_\omega H \} \setminus \{ z \}$ 
     for all $\omega'\in(x\to y)_{opt}$. 
     \end{itemize}
See Figure \ref{fig:saddles_essential} for an example of essential and unessential saddles.

Another important notion for metastability is the \emph{mixing time}. The mixing time of a Markov chain is the time needed for the law of the chain to be close to stationary.
More precisely, for every $0 < \epsilon < 1$, the mixing time $t^{mix}_\beta(\epsilon)$ is
\begin{equation}\label{eq:mixing_time_metropolis}
t^{mix}_\beta(\epsilon):=\min \{n \geq 0 \, | \, \max_{x \in \mathcal{X}} ||p^n_\beta(x, \cdot)-\mu (\cdot) ||_{TV} \leq \epsilon \},
\end{equation}
where $||\nu-\nu'||_{TV}:=\frac{1}{2}\sum_{x\in \mathcal{X}} |\nu(x)-\nu'(x)|$ for two probability distributions $\nu,\nu'$ on $\mathcal{X}$.

The mixing time is intimately connected to the \emph{spectral gap}, which relates the energy difference between the ground state and first excited state of a system. In particular, we will see in the following section (Theorem \ref{thm:mixing_time_spectral_gap_Metropolis}) that the spectral gap is equal to the logarithm of the mixing time over $\beta$ in the limit $\beta \to \infty$. We formally define the spectral gap as follows.
Let $(p_\beta(x, y))_{x,y \in \mathcal{X}}$ be the transition matrix of the Markov chain. The spectral gap is defined as
\begin{equation}\label{eq:spectral_gap_metropolis}
\rho_\beta:=1-a_\beta^{(2)},
\end{equation}
where $1=a_\beta^{(1)} > a_\beta^{(2)} \geq \cdots \geq a_\beta^{(|\mathcal{X}|)} \geq -1$ are the eigenvalues of the transition matrix.

\subsection{Metastability: main results}\label{metastability_main_results}
An important ingredient to study the transition time from the set of metastable states to the set of stable states is the so called \emph{recurrence property} of the Markov chain. In words, with probability super-exponentially close to one, starting from any state of $\mathcal{X}$ the process visits $\mathcal{X}_V$  within a time of order $e^{\beta V}$. 
    \begin{theorem}[Recurrence property]\cite[Theorem 3.1]{manzo2004essential}
    \label{thm:recurrence_property_Metropolis}
    For any $\epsilon>0$ and sufficiently large $\beta$ the following function is is SES\footnote{A function $\beta\to f(\beta)$ is called \textbf{super-exponentially small} 
(SES) if \begin{equation}\label{eq:ses_function} 
\lim_{\beta\to\infty}{\frac{1}{\beta}}\log f(\beta)=-\infty. 
\end{equation} },
\begin{equation}\label{eq:recurrence_property_SES}
        \beta \mapsto \sup_{\sigma\in \mathcal{X}} \mathbb{P}_\sigma\Big (\tau_{\mathcal{X}_V} > e^{\beta (V + \epsilon)} \Big ) .
      \end{equation}
    \end{theorem}
Equation~\eqref{eq:recurrence_property_SES} implies that the system reaches with high probability
a state in $\mathcal{X}_V$ in a time shorter than $e^{\beta (V+\epsilon)}$, uniformly in the starting configuration $\sigma$ for any $\epsilon >0$. In other words we can say that the dynamics speeded up by a 
time factor of order $e^{\beta V}$ reaches with high probability $\mathcal{X}_V$.

We note that the recurrence property is essential for the study of the tunnelling problem between stable states. 
Indeed, thanks to this property, it is possible to obtain results on $\tau^{\sigma}_{\mathcal{X}^s}$ in probability (Theorem \ref{thm:asymptotic_result_Metropolis_1}) and on the asymptotics of the expectation (Theorem \ref{thm:asymptotic_result_Metropolis_3}). 

   \begin{theorem}[Transition time] \cite[Theorem 4.1]{manzo2004essential} 
   \label{thm:asymptotic_result_Metropolis_1} 
   Let $\sigma_0 \in \mathcal{X}^m$ 
   and $V_{\sigma_0}=\Gamma$. 
   Then, for every $\epsilon>0$
    \begin{equation}
        \lim_{\beta \to \infty} \mathbb{P}_{\sigma_0} \Big ( e^{\beta(\Gamma-\epsilon)}< \tau_{\mathcal{X}^s} < e^{\beta(\Gamma+\epsilon)} \Big )=1.
    \end{equation}
   \end{theorem}
   
   \begin{theorem}[Expected value for the transition time] \cite[Theorem 4.9]{manzo2004essential}
   \label{thm:asymptotic_result_Metropolis_3}  
    Given $\sigma_0\in\mathcal{X}^m$, 
    \begin{equation}\label{eq:thm3_asymptotic_Metropolis}
     \lim_{\beta\to\infty}{\frac{1}{\beta}}\log  \mathbb{E}_{\sigma_0}[\tau_{\mathcal{X}^s}]=\Gamma.
    \end{equation}
   \end{theorem}

A strong version of Theorem \ref{thm:asymptotic_result_Metropolis_3} is the following sharp estimate of the expected value.

\begin{theorem}[Sharp estimate for the transition time] \cite[Theorem 16.5]{bovier2016metastability}
\label{thm:mean_crossover_time}
 Given $m\in\mathcal{X}^m$ and $s\in \mathcal{X}^s$, there exists a constant $K \in (0,\infty)$ such that
 \begin{equation}
     \lim_{\beta \to \infty} e^{-\beta \Gamma}\mathbb{E}_{m}(\tau_s)=K.
 \end{equation}
\end{theorem}
\noindent
This result holds under the assumption that the metastable and the stable states are unique. For the computation of $K$, a key role is played by the \textit{Dirichlet form} associated with a reversible Markov chain and by the \textit{capacity} as the solution of the Dirichlet principle, see \cite[Section 16]{bovier2016metastability} for details.

Next, we present a result that connects the notion of mixing time and the spectral gap with that of the energy barrier of the system.
\begin{theorem}[Mixing time and spectral gap for Metropolis Markov chains] \cite[Proposition 3.24]{nardi2016hitting} 
\label{thm:mixing_time_spectral_gap_Metropolis}
For any $0<\epsilon<1$ and any $s \in \mathcal{X}^s$,
\begin{equation}\label{eq:mixing_time_metropolis_result}
 \lim_{\beta \rightarrow \infty}{\frac{1}{\beta}\log{ t^{mix}_\beta(\epsilon)}}=\lim_{\beta \rightarrow \infty}-{\frac{1}{\beta}}\log\rho_{\beta}.
\end{equation}
Furthermore, there exist two constants $0<c_1<c_2<\infty$ independent of $\beta$ such that for every $\beta>0$,
\begin{equation}\label{eq:estimate_ro_metropolis}
 c_1e^{-\beta \, (\Gamma+\gamma_1) } \leq \rho_{\beta} \leq c_2e^{-\beta \, (\Gamma-\gamma_2) }.
\end{equation}
 where $\gamma_1,\gamma_2$ are functions of $\beta$ that vanish for $\beta\to\infty$.
\end{theorem}


\subsection{Clusters and Peierls contours}\label{sec:polyominos}

We introduce the notion of \emph{cluster} and we rewrite the energy \eqref{hamiltonianFunction} of a configuration $\sigma$ in terms of the latter. In this way, we can associate a pure geometrical figure, the \emph{polyomino}, to each cluster and find the shape of the critical configurations by solving an isoperimetric problem on the perimeter and the area of the polyominos.

\begin{figure}[htb!]
\centering
\includegraphics[scale=0.8]{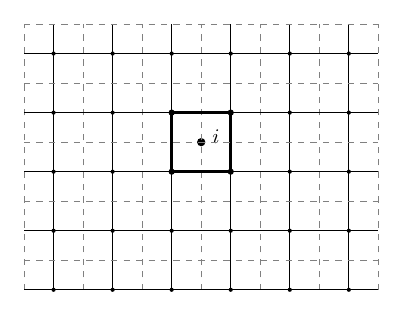} \,\,\,\,\,\,\,\,
\includegraphics[scale=0.8]{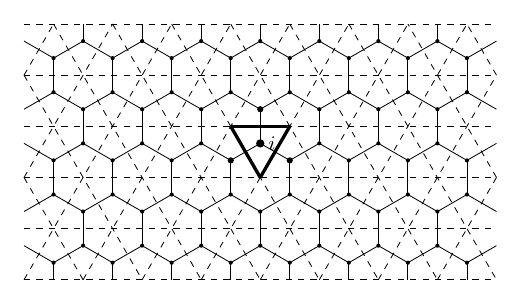} 
    \caption{The solid lines show the lattice $\mathbb{L}$,
            whereas the dashed lines show its dual, $\mathbb{D}$. The solid square on the left and the solid triangle on the right highlight a face of $\mathbb{D}$ centered
            at site $i \in \mathbb{L}$. The thicker vertices are the
            nearest neighbors of site $i$ on the 
            $\mathbb{L}$.
    }
  \label{neighbors}
    \end{figure}

Given a lattice $\mathbb{L}$, the \emph{face} or \emph{cell} is the subset of $\mathbb{R}^2$ centered at site $x \in \mathbb{L}$ that, when repeated in a regular pattern, forms the entire lattice structure. Two sites $i,j \in \mathbb{L}$, belonging to the same cell, are \emph{nearest neighbors} when they share an edge of the lattice.
The dual lattice $\mathbb{D}$ of $\mathbb{L}$ is a lattice such that each of its vertices is the center of each cell of $\mathbb{L}$, and
every of its pairs of vertices is connected if the corresponding cells of $\mathbb{L}$ share an edge.
In particular, if $\mathbb{L}$ is a square lattice, then $\mathbb{D}$ is still a square lattice and a cell is a unit square, see the left panel of Figure \ref{neighbors}. Otherwise if $\mathbb{L}$ is a hexagonal lattice, then $\mathbb{D}$ is a triangular lattice, and a cell of $\mathbb{L}$ is a unit hexagon while the one of $\mathbb{D}$ is a unit triangle, see the right panel of Figure \ref{neighbors}.

Given a configuration $\sigma \in \mathcal{X}$, consider the set $\mathcal{C}(\sigma) \subset \mathbb{R}^2$ defined as the union of the closed unit faces of $\mathbb{D}$ centered at sites $x\in \mathbb{L}$ such that $\sigma(x)=1$. The maximal connected components $C_1 ,..., C_m$, $m\in \mathbb{N}$, of $\mathcal{C}(\sigma)$ are called \emph{clusters} of $\sigma$.

Given a configuration $\sigma \in \mathcal{X}$ we denote by $\gamma(\sigma)$ its Peierls contour that is the boundary of the clusters. 
Note that the Peierls contours live on the dual lattice and are the union of piecewise linear curves separating spins with opposite sign in $\sigma$. In particular, if $\mathbb{D}$ is a square lattice then in each dual vertex there are 0, 2, 4 dual bonds contained in $\gamma(\sigma)$, instead if $\mathbb{D}$ is a triangular lattice than in each dual vertex there are 0, 2, 4, 6 dual bonds contained in $\gamma(\sigma)$.

We can rewrite the Hamiltonian function in terms of Peierls contours and number of plus spins $N^{+}(\sigma)=\sum_{x \in \Lambda}\frac{\sigma(x)+1}{2}$ in the following way,
\begin{align}\label{eq:peierls_hamiltonian}
    H(\sigma)-H(\textbf{-1})=J|\gamma(\sigma)|-h N^{+}(\sigma).
\end{align}

So the energy of each configuration is associated to the area and the length of the boundary (the perimeter) of a suitable collection of unit faces.
Thanks to \eqref{eq:peierls_hamiltonian} it is possible not only to find the shape of the critical configurations and the value of the energy barrier (that is the energy of these configurations), but also to compute a sharp estimate of the transition time including the prefactor as in Theorem \ref{thm:mean_crossover_time}. Indeed, as we will see in the next two sections, the prefactor depends on the \emph{critical length} of the critical droplets.

\subsection{Square lattice}\label{subsec:square}

\begin{figure}[htb!]
\centering
    \includegraphics[width=\textwidth, scale=0.85]{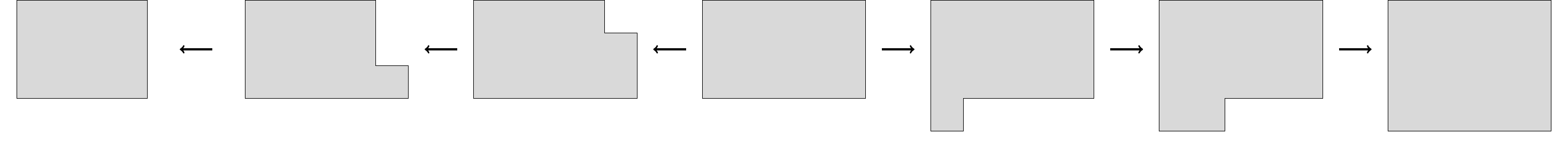}
    \caption{From the central panel to the left, the rectangle shrinks. From the central panel to the right, the rectangle grows.}
  \label{Ising_evolution_2D}
    \end{figure}

The earliest studies of metastability at low temperatures for the Ising model on a two-dimensional square lattice date back to the two works \cite{neves1991critical, neves1992behavior}. In \cite{neves1991critical}, the authors use the pathwise approach to investigate the metastable behavior of the Ising model on a finite 2D torus evolving under Glauber dynamics defined as a continuous-time Markov chain  where the rate of a spin at the site $x\in \Lambda$ is equal to $$c(x,\sigma)=\exp\{-\beta [H(\sigma^{(x)})-H(\sigma)]_+\},$$ with $\sigma^{(x)}$ defined as in \eqref{sigma_x}.
For the first time they prove that, assuming a positive external magnetic field $h>0$ small enough, the rectangular droplet of pluses in a sea of minuses either shrinks or grows depending on its size, which is a function of $h$, see Figure \ref{Ising_evolution_2D} to understand how.

\begin{figure}[htb!]
\centering
    \includegraphics[scale=0.8]{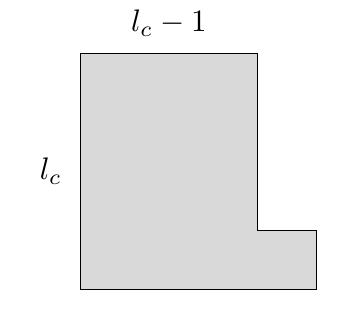}
    \caption{An example of a critical droplet for the Ising model on the 2D torus.}
  \label{Ising_droplet}
    \end{figure}

Specifically, without loss of generality, they assume $J=1$ and find that a rectangle with side length greater than $l_c=\lceil 2/h \rceil$ ($l_c=\lceil \frac{2J}{h}\rceil$ for general $J>0$) tends to grow, otherwise it tends to shrink.
Consequently, starting from the homogeneous state $\textbf{-1}$ and moving toward the stable state $\textbf{+1}$, the system crosses critical configuration containing a \emph{quasi-square} $l_c \times (l_c-1)$ of pluses with a protuberance attached along one of the longer sides in a sea of minuses, see Figure \ref{Ising_droplet}.

As a result, starting from the homogeneous configuration $\textbf{-1}$, the system stays close to this state for an unpredictable time until a critical square droplet of a precise size is formed, then it reaches the stable state $\textbf{+1}$ in a relatively short time. They find that the asymptotic value for the time of the total decay is related to the height of an energy barrier, that corresponds to the energy of the critical droplet (Theorems \ref{thm:asymptotic_result_Metropolis_1}, \ref{thm:asymptotic_result_Metropolis_3}), which is equal to
\begin{align}
\Gamma=4l_c-(l_c^2-l_c+1)h.
\end{align}
In \cite{neves1992behavior}, the authors extend the results obtained in the previous paper to encompass a class of Glauber dynamics for the 2D nearest neighbor ferromagnetic Ising model, in which the spin-flip depends only on the energy increment caused by the flip following a monotonic non-increasing function. For instance, in addition to the Glauber dynamics defined above,
other cases may include the heat bath dynamics with rate 
$c(x,\sigma)=(1+\exp\{\beta [H(\sigma^{(x)})-H(\sigma)]_+\})^{-1}$, 
and another Glauber dynamics with rate $c(x,\sigma)=\exp\{-\beta [H(\sigma^{(x)})-H(\sigma)]_+ /2\}$.

The author of \cite{schonmann1992pattern} analyzes the same model studied in \cite{neves1991critical}, but under the assumption that Glauber dynamics defined as a discrete-time Markov chain as in \eqref{def:glauber}. Since the law of large numbers for the Poisson process enables to translate easily results from the continuous time version into the discrete time version and back, then all the results in \cite{neves1991critical} remain valid for the discrete time case with minor adjustments to the proofs.
In particular, \cite{schonmann1992pattern} offers a detailed explanation of how the system reaches the stable state $\textbf{+1}$: starting from $\textbf{-1}$, the system through a sequence of configurations featuring growing clusters that are \emph{as close as possible to quasi-square} in shape.

The study of the metastability for the 2D Ising model on the torus is concluded in 
\cite{boviermanzo2002metastability}, where a sharp estimate of the transition time is provided through the computation of the prefactor (Theorem \ref{thm:mean_crossover_time}), i.e.,
\begin{align}
    \lim_{\beta \to \infty} e^{-\beta \Gamma}\mathbb{E}_m(\tau_s)=\frac{3}{8}\frac{1}{l_c-1}.
\end{align}

Another of the first analyses of the metastable behavior of the Ising model is \cite{dehghanpour1997metropolis}. There, the authors consider the model on infinite volume $\mathbb{Z}^2$ and show that, at low temperature and under Glauber dynamics, the relaxation time is of the order $\exp{\beta \mathcal{K}}$ with $\mathcal{K}=\frac{\Gamma-(2-h)}{3}$, where the factor $1/3$ derives from the growth and shrinkage mechanism of the droplet and it is related to the dimension of the lattice, while the term $(2-h)$ is related to the rate of growth of highly supercritical droplets. An additional study of the metastability in large volumes is presented in \cite{bovier2010homogeneous}, where the authors study the evolution of the Ising model under Glauber and Kawasaki dynamics on a square box $\Lambda_\beta \subset \mathbb{Z}^2$ with periodic boundary conditions such that $\lim_{\beta \to \infty} |\Lambda_\beta|=\infty$.

Furthermore, another important contribution is found in \cite{manzo2004essential}, which presents a straightforward strategy to the study of metastability for general Metropolis Markov chains, with a particular focus on applications to stochastic dynamics in lattice spin systems.
The approach involves decoupling the asymptotic behavior of the transition time from the characterization of the tube of typical paths realizing the transition. This approach proves particularly valuable when determining the tube of typical paths is too difficult. Additionally, they analyze the structure of the saddles introducing the notion of \emph{essential saddles} as we reported in Section \ref{metastability_main_tools}. Finally, to illustrate their methodology, they apply it to the case of Glauber dynamics for 2D Ising model also in the degenerate case, i.e. assuming $\frac{2J}{h}$ is integer. In particular, they find that the set of the critical configurations is composed by all configurations containing a rectangle $l_c \times m$ with $l_c=\frac{2J}{h}$ and $m=l_c+1,...,L-1$, having a protuberance attached to one of its sides. This result is different in the non-degenerate case ($\frac{2J}{h} \not \in \mathbb{N}$), in which there is only one minimal gate, that is the quasi-square of pluses $l_c \times (l_c-1)$ with a protuberance attached along the longest side.

Different results of the metastable behavior of the Ising model on square lattice are obtained in \cite{cirillo1998metastability}, where the authors assume free boundary conditions. In this case, the nucleation is not homogeneous, indeed they show that the exit from the metastable phase $\textbf{-1}$ occurs via the formation of a critical square droplet of pluses in one of the four corners of $\Lambda$. The tendency of the droplet to grow is favored when such a droplet has one of its sides on the boundary, this implies that 
the exit time is much smaller than in the case of periodic boundary conditions, see Table \ref{tab:free_periodic} for for further details.

\begin{table}[H]
\begin{center}
\small{
\begin{tabular}{c|c|c|c}
\hline\hline
& critical length & energy barrier $\Gamma$ & exit time $\tau$ \\
\hline\hline
periodic b.c. & $\lceil 2J/h\rceil$ & $4J \lceil 2J/h\rceil-( \lceil 2J/h\rceil^2-\lceil 2J/h\rceil+1)h$ & $\tau \sim \exp\{\beta(4J^2 /h)\}$ \\
\hline
free b.c. & $\lceil J/h\rceil$ & $4J \lceil J/h\rceil-(\lceil J/h\rceil^2-\lceil J/h\rceil+1)h$ & $\tau \sim \exp\{\beta(J^2 /h)\}$ \\
\hline\hline
\end{tabular}
}
\end{center}
\caption{On the first row, the value of the critical length, the energy barrier and the exit time in the case of periodic boundary conditions. On the second row, the values in case of free boundary conditions.
}
\label{tab:free_periodic}
\end{table}

\begin{figure}[htb!]
\centering
    \includegraphics[scale=0.3]{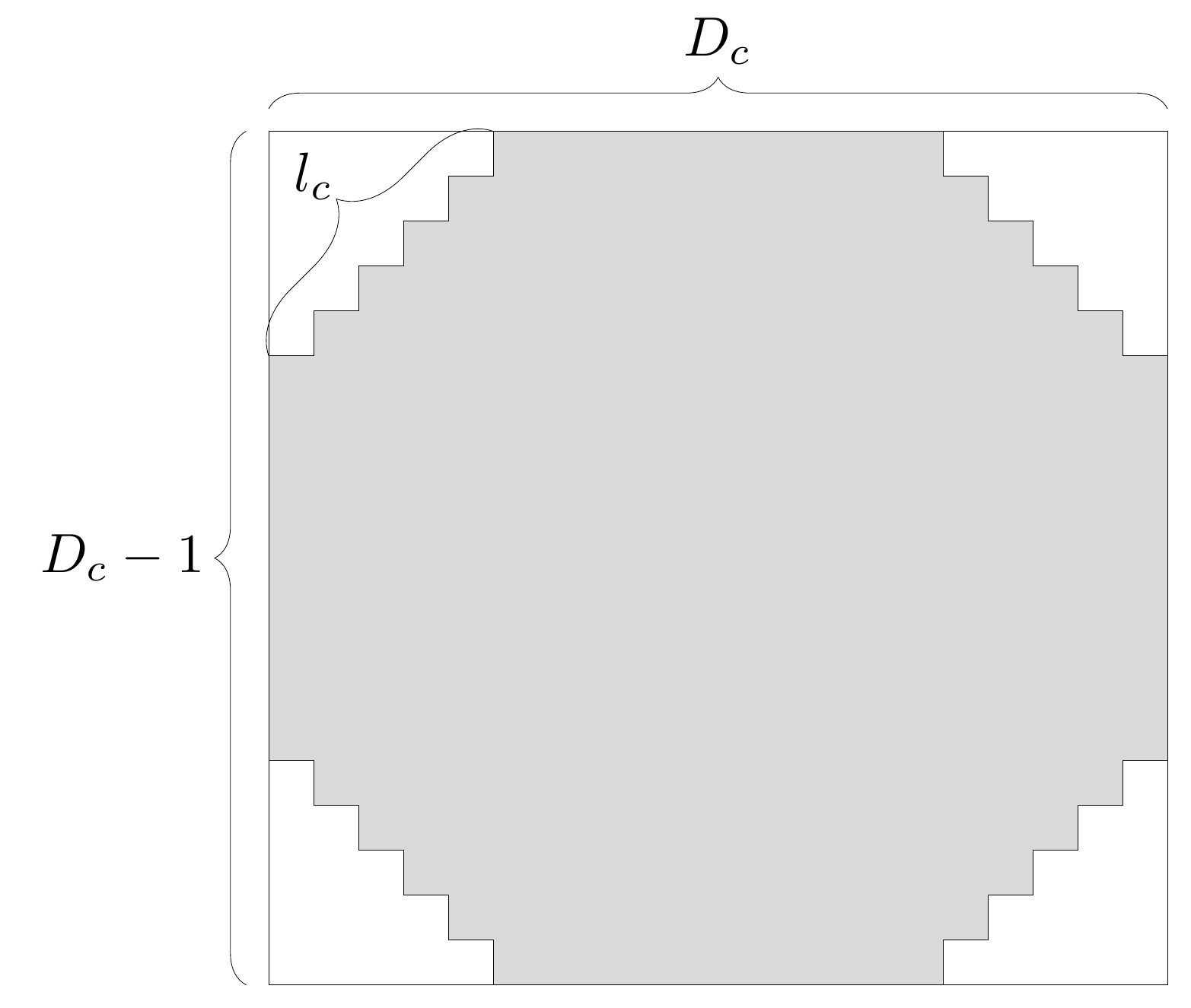}
    \caption{An example of a critical droplet for the Ising model with nearest neighbors and next neighbor interaction.}
  \label{Ising_next_neigh}
    \end{figure}

In \cite{kotecky1994shapes}, the authors consider a 2D Ising model with nearest neighbors and next neighbor interaction. They study the metastable behavior of this model under Glauber dynamics defined in \eqref{def:glauber} assuming periodic boundary conditions. They define the following Hamiltonian of the system
\begin{align}
   H(\sigma):=-\frac{J}{2}\sum_{\substack{i,j \in \Lambda\\ d(i, j)=1}} \sigma (i) \sigma (j)
   -\frac{K}{2}\sum_{\substack{i,j \in \Lambda\\ d(i, j)=\sqrt{2}}} \sigma (i) \sigma (j)-\frac{h}{2} \sum_{i \in \Lambda} \sigma (i),
\end{align}
characterize the shape of the critical configurations and present a detailed description of the escape pattern in the asymptotic region of vanishing temperatures. In particular, the critical droplets are octagons with oblique side length $l_c=\left \lceil \frac{2K}{h} \right \rceil$ inscribed in a quasi-square $D_c \times (D_c-1)$ where $D_c=\left \lceil \frac{2J}{h} \right \rceil$, and a protuberance attached, see Figure \ref{Ising_next_neigh}.

\subsection{Hexagonal lattice}\label{subsec:hexagonal}
In \cite{apollonio2021metastability}, the authors consider the Ising model on the hexagonal lattice evolving with Glauber dynamics.  
In particular, they consider $\Lambda$ be the subset of the hexagonal lattice obtained by
cutting a parallelogram of side length $L$ along two of the coordinate
axes of the triangular lattice. 

\begin{figure}[htb!]
\centering
    \includegraphics[width=\textwidth, scale=0.9]{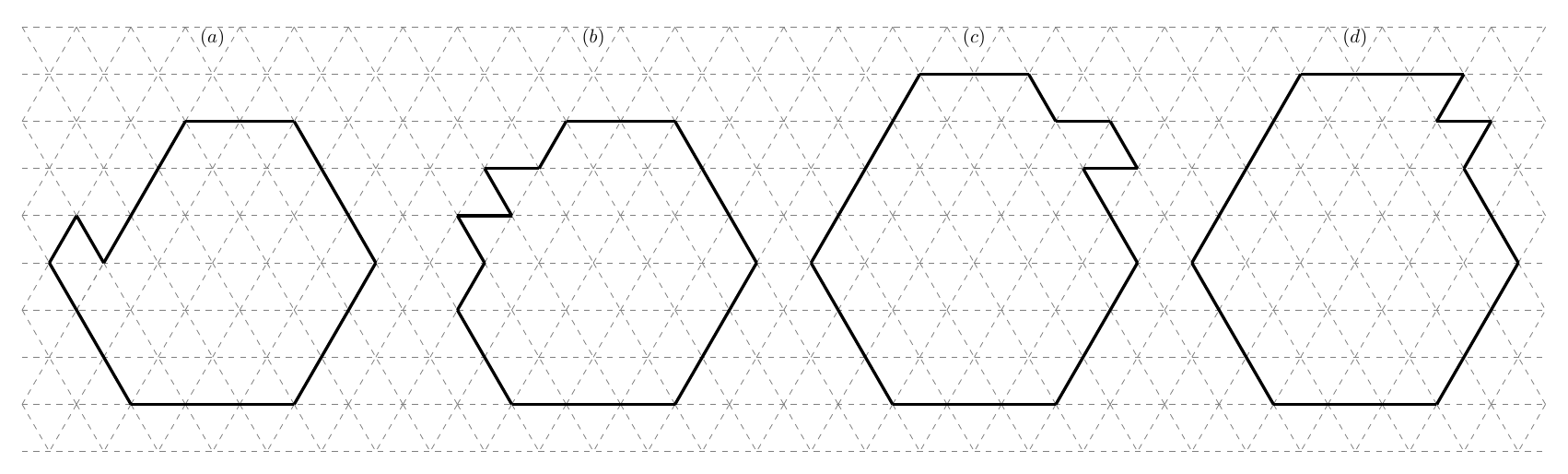}
    \caption{On the left there are two examples of two cluster shapes belonging to the gate for $\delta \in (0,1/2)$. On the right there are other two examples of two cluster shapes for $\delta \in (1/2,1)$.}
  \label{figselle}
    \end{figure}

The external magnetic field $h$ and the ferromagnetic interaction $J$ in the Hamiltonian function \eqref{hamiltonianFunction} 
are chosen such that $h \in (0,1)$ and $J >> h$, so that the system can exhibit a metastable behavior.
To ensure the non degenerate behavior of the Ising model, the authors assume $\frac{J}{2h}-\frac{1}{2} \not \in \mathbb{N}$. 
Under the assumption that the finite torus is large compared to the size
of the critical clusters, that is $|\Lambda| \geq (\frac{4J}{h})^2$,
the authors prove that $\textbf{-1}$ is the unique metastable state of the system and they compute the energy barrier to reach $\textbf{+1}$ starting from $\textbf{-1}$. To do this, first they show that the only two states with stability level strictly greater than $2J$ are $\textbf{-1}$ and $\textbf{+1}$, i.e. $\mathcal{X}_{2J}=\{ \textbf{-1}, \textbf{+1} \}$, and then by using Theorem \ref{thm:recurrence_property_Metropolis} they prove that for any $\epsilon>0$, the following function is SES,
\begin{align}\label{recurrence2J}
    \beta \mapsto \sup_{\sigma \in \mathcal{X}} \mathbb{P}_{\sigma}(\tau_{\mathcal{X}_{2J}}> e^{\beta(2J+\epsilon)}).
\end{align}
They prove that the critical configurations contains a cluster having a shape
that is close to a hexagon with area equal to
\begin{equation}\label{areacritica}
\begin{cases}
        A^*_1= 6{r^*}^2+10r^*+5 & \text{if } 0<\delta<\frac{1}{2}, \notag \\
        A^*_2= 6(r^*+1)^2+2(r^*+1)+1 & \text{if } \frac{1}{2}<\delta<1.
\end{cases}
\end{equation}
where $r^*$ is the \emph{critical radius} such that 
\begin{equation}\label{raggiocritico}
    r^*:=\left\lfloor \frac{J}{2h}-\frac{1}{2} \right\rfloor, 
\end{equation}
and $\delta \in (0,1)$ is the fractional part of $\frac{J}{2h}-\frac{1}{2}$, that is $\delta = \frac{J}{2h}-\frac{1}{2} - r^*$.
See Figure \ref{figselle}.

In other words, given $\delta \in (0,1)$ and $A^*_i \in \{A^*_1, A^*_2 \}$ in \eqref{areacritica}, the configuration containing cluster as in  Figure \ref{figselle} is the union of all minimal gates from the homogeneous state $\textbf{-1}$ to the homogeneous one $\textbf{+1}$. 

The value of the maximal stability level $\Gamma^{Hex}$ is obtained by 
computing the energy of the critical configurations, and it is equal to
\begin{equation}\label{GammaH}
    \Gamma^{Hex}:=
    \begin{cases}
        -6{r^*}^2h+6r^*J-10r^*h+7J-5h & \text{if } 0<\delta<\frac{1}{2} \\
        -6(r^*+1)^2h+6(r^*+1)J-2(r^*+1)h+3J-h & \text{if } \frac{1}{2}<\delta<1
    \end{cases}
\end{equation}

Using the value of the energy barrier, they obtain the same results present in Theorems \ref{thm:asymptotic_result_Metropolis_1} , \ref{thm:asymptotic_result_Metropolis_3} , \ref{thm:mixing_time_spectral_gap_Metropolis} about the transition time, the mixing time and the spectral gap. 

After the identification of the critical configurations according to their shape and size, and after to compute the energy of them, it is possible to find a sharp estimate of the expected value of the transition time. In particular, they find
\begin{equation}\label{valoreatteso}
     \mathbb{E}_{\textbf{-1}}[\tau_{\textbf{+1}}]=\frac{1}{K}e^{\beta\Gamma^{Hex}}(1+o(1)), \,\,\, \text{  where } \,\,\, K:=
\begin{cases}
5(r^*+1) & \qquad \text{if } \delta \in (0,\frac{1}{2}), \\
10(r^*+1) & \qquad \text{if } \delta \in (\frac{1}{2},1).
\end{cases}
\end{equation}

\subsection{Graphs and random graphs}\label{subsec:random_graph}

The study of metastability on graphs highlights the complex behavior and difficulties in analyzing systems with \emph{non-uniform} or \emph{random} connections, where both thermal fluctuations and the (random) structure of the graph prevent the system from quickly reaching a globally stable state. In this Section, we will report interesting results on the metastable behavior of the Ising model on graphs and random graph. 

In \cite{dommers2017metastability}, the author studies the metastability of the ferromagnetic Ising model on a random $r$-regular graph in the zero temperature limit with a small external magnetic field. Fixed the number of vertices $n$ large enough, he estimates that during the transition from the homogeneous state $\textbf{-1}$ to the homogeneous state $\textbf{+1}$ the system overcomes an energy barrier $\Gamma$ such that 
\begin{align}
    (r/2-C_1 \sqrt{r})n \leq \Gamma \leq (r/2+C_2 \sqrt{r})n
\end{align}
where $C_1,C_2$ are two positive constants.
Thus, he states that the transition time between these two states behaves like $e^{\beta n(r/2 + O(r))}$.

In \cite{dommersnardi2017metastability}, the authors obtain a similar estimate of the energy barrier when they consider the behavior of the Ising model on a random multi-graph known as \emph{configuration model}.

The metastable behavior of the Ising model on random graphs has also been analyzed for the \emph{Erdős–Rényi random graph} with a fixed edge retention probability in \cite{hollander2021glauber} and \cite{bovier2021metastability}. Both studies compare the mean metastable exit times of the random model to those of the standard Curie–Weiss model in large volumes at a fixed temperature. 
In particular, in \cite{hollander2021glauber} a pathwise approach demonstrates that the mean metastable exit times are asymptotically equivalent to those of the Curie–Weiss model, scaled by a random prefactor that depends on the system size and has polynomial order. The potential-theoretic approach in \cite{bovier2021metastability} improves the estimate of the prefactor but at the cost of reduced generality in the choice of the initial distribution. 

The results for the Erdős–Rényi random graph are further extended in \cite{bovierslowik2022metastability} to encompass inhomogeneous dense random graphs and more general random interactions. In this work, the authors compare the metastable behavior of a class of spin systems whose Hamiltonian has random and conditionally independent coupling coefficients, called quenched model, with the corresponding annealed model in which the coupling coefficients are replaced by their conditional mean.

\cite{bovier2022metastability} contains an investigation of metastability under Glauber dynamics for the Ising model on the \emph{complete graph with random independent couplings} in large volumes and at a fixed temperature. The authors of this work obtain sharp estimates on mean metastable exit times using the potential-theoretic approach with coarse-graining techniques. 

\begin{figure}[htb!]
 \centering
     \includegraphics[scale=0.3]{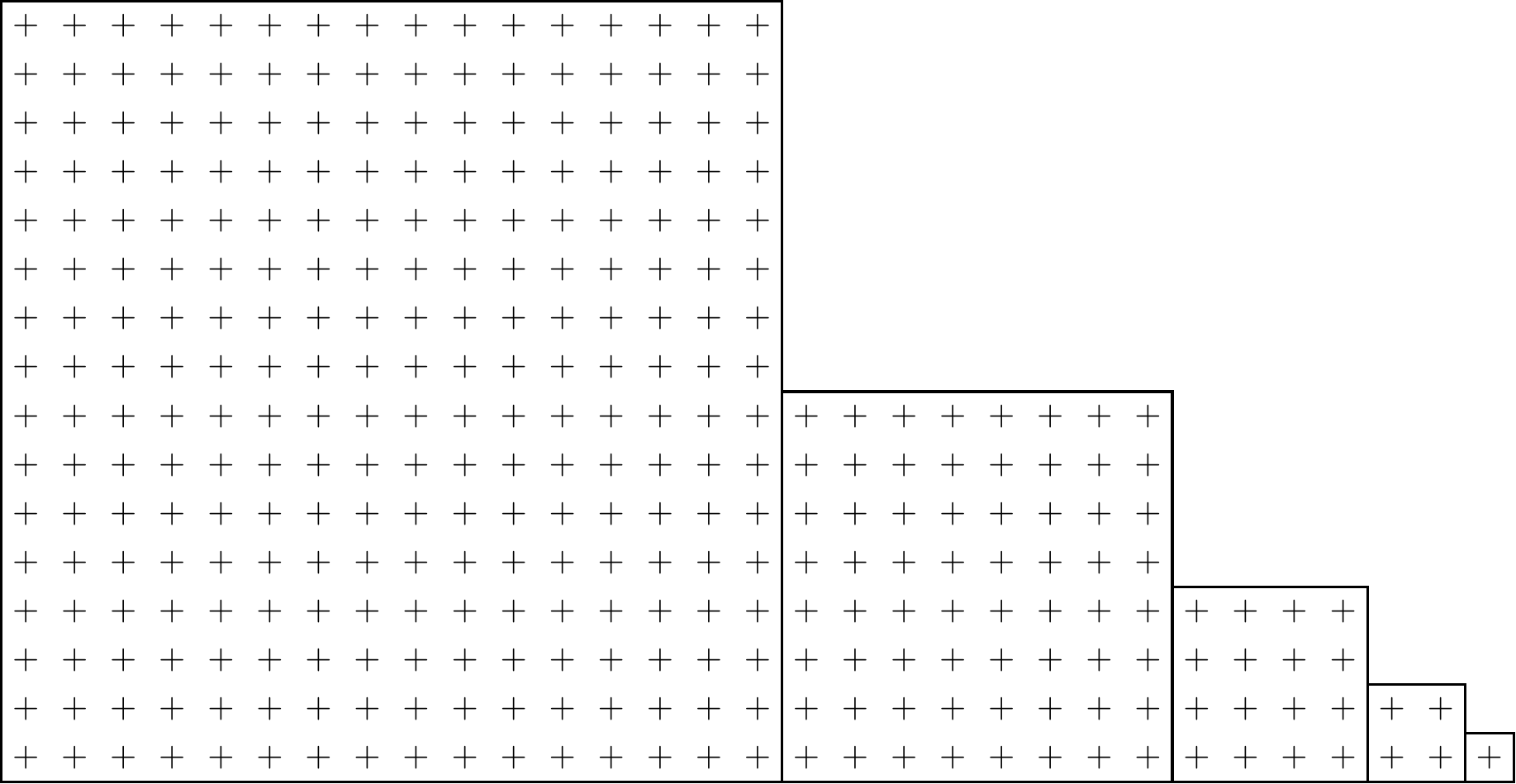} 
    \caption{An example of the critical droplets in \cite{jovanovski2017metastability}.}
  \label{fig_iper}
    \end{figure}

Another interesting result on the asymptotic behavior of the exit time is present in \cite{jovanovski2017metastability}. In this paper, the author studies the metastability for the Ising model on a $n$-dimensional hypercube and he uses potential theoretic approach to derive a sharp estimate of the prefactor (Theorem \ref{thm:mean_crossover_time}). In particular, 
for small value of the external magnetic field $0<h<n-2$, he computes the energy barrier from the detected metastable state $\textbf{-1}$ and the stable one $\textbf{+1}$, and he finds the value of the prefactor for the estimate of the transition time:
\begin{align}
  \Gamma:= \frac{1}{3} (2-h+ \lfloor h \rfloor ) \left ( 2^{\lceil n-h \rceil} + 2 \varepsilon -1 \right ) - \varepsilon \qquad \text{and}
  \qquad K:=\frac{3\lceil h \rceil !}{ (n! 2^n (1+\varepsilon))},
\end{align}
where $\varepsilon=\lceil n-h \rceil \text{mod}2$. 
Moreover, he identifies the shape of the critical droplets as in Figure \ref{fig_iper}. Roughly speaking, these configurations contain a sequence of attached squares of pluses with side length $2^{i-1}$ for $i=1,...,\lceil (n-h)/2\rceil $.

\begin{figure}[htb!]
\centering
    \includegraphics[scale=0.4]{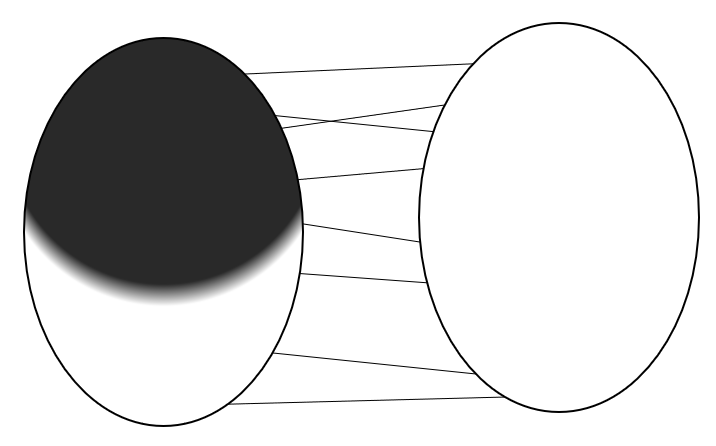} \,\,\,\,\,\,\,\,\, \includegraphics[scale=0.4]{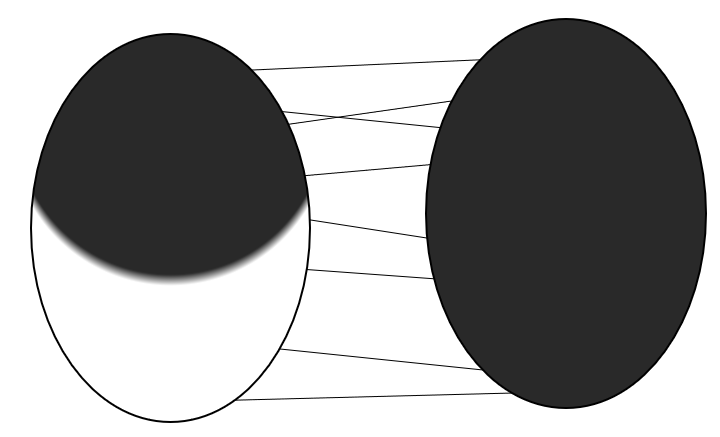}
    \caption{An example of the critical configurations, where the black color denotes the plus spins and the white color indicates the minus spins.
    }
  \label{fig_uova}
    \end{figure}

Recently, in \cite{baldassarri2023ising} a model for opinion dynamics has been developed considering the Ising model with an external magnetic field $h \geq 0$ on a family of finite networks with a clustered structure. In particular, the authors consider the Ising model on a graph consisting of two clusters of equal size, which are locally complete graphs, and such that each node is connected to a single node in each of the other clusters. Rigorous estimates in probability, expectation, and law for the first hitting time between metastable (or stable) states and (other) stable states are derived at low-temperature regime, in which homogeneous opinion patterns prevail and, as such, it takes the network a long time to fully change opinion (the analogous of the Theorems \ref{thm:asymptotic_result_Metropolis_1}, \ref{thm:asymptotic_result_Metropolis_3}). Moreover, the authors find tight bounds on the mixing time and spectral gap of the Markov chain (i.e. Theorem \ref{thm:mixing_time_spectral_gap_Metropolis}) and they characterize the critical configurations for the dynamics, see Figure \ref{fig_uova} for instance.

\section{Ising Model in higher dimensions}\label{sec:3D}
In this Section, we will show some results on the metastable behavior of the Ising model in $d$-dimensions, with $d\geq 3$. 

\begin{figure}[htb!]
\centering
    \includegraphics[scale=0.4]{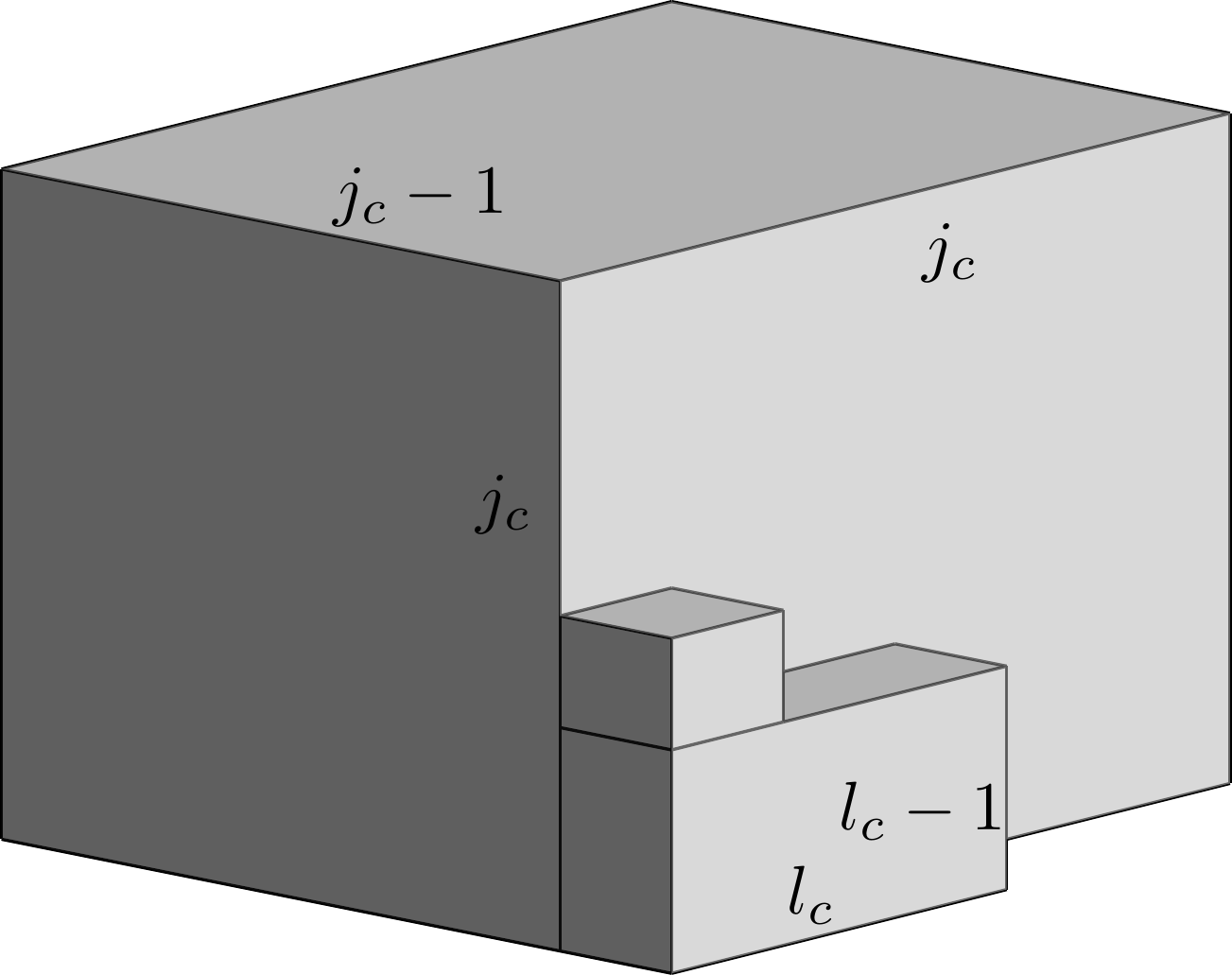}
    \caption{An example of the critical droplet for 3D Ising model.
    }
  \label{3d_Ising}
    \end{figure}

The 3D Ising model is a generalization of the classic Ising model and it is much more difficult to analyze due to the increased number of interacting neighbors and the greater geometric complexity. In \cite{arous1996metastability} the authors study this model on a 3D finite torus and they analyze the evolution of the system at low temperature, introducing Glauber dynamics. To compute the energy barrier that the system must overcome to reach the stable state, they use Freidlin-Wentzell theory, which divides the state space into cycles based. Indeed, these cycles form a partition of the configuration space based on energy levels and they represent the configurations that the system explores before transitioning to a stable state. 
However, in higher dimension the energy landscape becomes significantly more complex than in two-dimensional cases, so the authors of \cite{arous1996metastability} develop some methods to deal with this complexity and analyze the nucleation and the metastable behavior. They identify the critical configurations and the tube of trajectories extending previous results for the 2D cases, see also \cite{alonso1996three} for the geometrical results. 
In particular, setting $J=1$ and recalling the definition of the Hamiltonian function \eqref{hamiltonianFunction} where the box has three-dimensions, they find a unique metastable state $\textbf{-1}$ and a stable state $\textbf{+1}$, and they prove that the value of the energy barrier is the energy of a critical droplet with volume equal to $j_c (j_c-1)(j_c-\delta_c)+l_c(l_c-1)+1$ where 
\begin{align}
    &l_c:= \left \lceil \frac{2}{h} \right \rceil \qquad \text{ is the 2D critical length}, \\
    &j_c:= \left \lceil \frac{4}{h} \right \rceil \qquad \text{ is the 3D critical length},
\end{align}
and $\delta_c$ is a parameter that depends on the external magnetic field and takes value $\delta_c=1$ if $4+ \sqrt{16 + h^2} < h \, (2\lceil 4/h \rceil-1)$ and $\delta_c=0$ otherwise.
We note that a similar parameter $\delta$ appears also in the case of 2D Ising model on the hexagonal lattice, see Section \ref{subsec:hexagonal}. The presence of $\delta_c$ arises from the geometry of the lattice and the requirement for $h$ to take a non-integer value, ensuring the system does not fall into a degenerate case.
These 3D critical droplets of pluses are parallelepipeds $j_c \times (j_c-1) \times (j_c-\delta_c)$ with attached a quasi-square $l_c \times (l_c-1)$ with a protuberance along the longest side, see Figure \ref{3d_Ising}.
For a complete description of the exit paths to reach the stable state see \cite{arous1996metastability}.

In two earlier articles \cite{neves1994stability, neves1995discrete}, the metastable behavior of an Ising model in $d$-dimensions is analyzed. 
In \cite{neves1995discrete}, using an induction on the dimension, the author proves the $d$–dimensional discrete isoperimetric inequality from which he deduces the asymptotic behavior of the relaxation time. The problem involves minimizing the surface area of subsets of the lattice $\mathbb{Z}^d$ for a fixed volume, where the surface area is defined by the number of edges with only one endpoint in the subset.
In dimensions $d>2$, the critical configurations are constructed using a hierarchical approach, where lower-dimensional blocks (e.g., critical droplets in dimension $d-1$) are combined along the lattice directions to preserve the minimization of surface area. 

Thanks to this construction of the critical droplets, the authors of \cite{cerf2013nucleation} 
show that the asymptotics of the relaxation time are of the order $e^{\beta \Gamma}$, where the energy barrier $\Gamma$ depends on the dimension $d$ of the lattice and on the external magnetic field $h$. In particular,
\begin{align}
    \Gamma=\frac{1}{d+1} (\Gamma_1+ ...+\Gamma_d),
\end{align}
where $\Gamma_i$ is the energy of the $i$-dimensional critical droplet of the Ising model at zero temperature and magnetic field $h$. 
Moreover, from the results in \cite{neves1994stability, neves1995discrete}, the authors of \cite{cerf2013nucleation} find that, let $l_c(d)= \left \lfloor \frac{2(d-1)}{h} \right \rfloor$, the critical configurations contains a quasi-cube $l_c(d) \times (l_c(d)+1)$ with a $(d-1)$-dimensional critical droplet attached on one of its largest sides. We observe that the precise shape of the critical droplets depends on the value $h$.

Referring back to \cite{arous1996metastability}, fixed $d=3$, the authors generalize the results of \cite{neves1994stability, neves1995discrete} by introducing the projection operators to reduce efficiently the polyominoes and to obtain the uniqueness of the minimal shapes for specific values of the volume. Furthermore, with a precise investigation of the energy landscape near these minimal shapes, they obtain full information on the exit path.

Finally, a sharp estimate of the transition time for $d$-dimensional Ising model on the torus is given in \cite{boviermanzo2002metastability}.

\section{Anisotropic Ising model}\label{sec:aniso}
The anisotropic Ising model is a generalization of the classical Ising model, where the energy of the system is no longer spatially uniform. This anisotropy can arise from an alternating external magnetic field applied across different regions of space, as in \cite{nardi1996low, nardi1999ising}, or from varying interaction strengths along different spatial directions, as in \cite{kotecky1993droplet}. The presence of the anisotropy offers a more accurate depiction of materials exhibiting structural or interaction-induced anisotropy and the flexibility of the model makes it a powerful framework for investigating complex phenomena in magnetic systems, critical transitions, and various anisotropic physical processes.

In \cite{kotecky1993droplet} the authors investigate the metastable behavior of an anisotropic Ising model, where varying interaction strengths along different directions are introduced to mimic the physical properties of specific real-world systems. They consider the following Hamiltoninan function defined on the torus assuming periodic boundary condition,
\begin{align}
    H(\sigma):=-\frac{J_1}{2}\sum_{\substack{i,j \in \mathcal{H}(\Lambda)}} \sigma (i) \sigma (j) 
    -\frac{J_2}{2}\sum_{\substack{i,j \in \mathcal{V}(\Lambda)}} \sigma (i) \sigma (j) -\frac{h}{2} \sum_{i \in \Lambda} \sigma (i),
\end{align}
where $\mathcal{H}(\Lambda)$ (resp. $\mathcal{V}(\Lambda)$) is the set of the horizontal (resp. vertical) nearest neighbor pairs of sites in $\Lambda$ and $J_1>J_2>>h>0$. They identify the metastable state as the homogeneous state $\mathbf{-1}$ and the stable state as $\textbf{+1}$ and they provides results on the shape of the critical droplets and on the transition time at low temperature (Theorem \ref{thm:recurrence_property_Metropolis}, \ref{thm:asymptotic_result_Metropolis_1}, \ref{thm:asymptotic_result_Metropolis_3}). In particular, they show that the critical droplet has not the shape of a polyomino with minimal perimeter fixed the area, indeed when the coupling constants are chosen such that $J_1 > J_2 > 0$ along the axes, the Wulff shape is a rectangle with a side length proportional to $J_1$ and the other one to $J_2$. Instead, in this case, the critical droplet has the shape of a square with side length $l^*=\lfloor2J_2/h\rfloor$. Going into details, the critical configuration is a square $l^* \times l^*$ with a protuberance attached along the vertical side, indeed this direction is energetically favorable since $J_1>J_2$ and the system moves in the direction with lower surface energy. Moreover, the authors describe the set of paths that the system follows with high probability to reach the stable state $\textbf{+1}$. This set contains many configurations that have not Wulff shapes. Indeed, starting from $\textbf{-1}$, the system reaches the critical droplet in an exponentially long time by forming subcritical configurations with rectangular shapes that
tend to shrink rather than grow in order to reach a state with lower energy.
After the formation of the critical droplet, the system continues to grow in a rectangular shape with growth favored in the vertical direction and, when one of the sides reaches its maximum size by wrapping around the torus, the growth extends in the horizontal direction.

\begin{figure}[hbt!]
    \centering
    \includegraphics[width=0.4\linewidth]{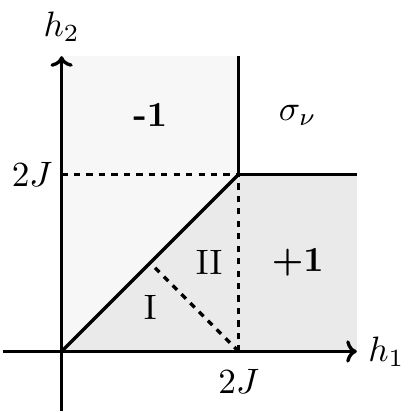}
    \caption{The zero-temperature phase diagram in the plane $h_1$-$h_2$. We reported the different stable states in different regions.}
    \label{fig:phase_diagram}
\end{figure}

\begin{figure}[htb!]
    \centering
    \includegraphics[width=1\linewidth]{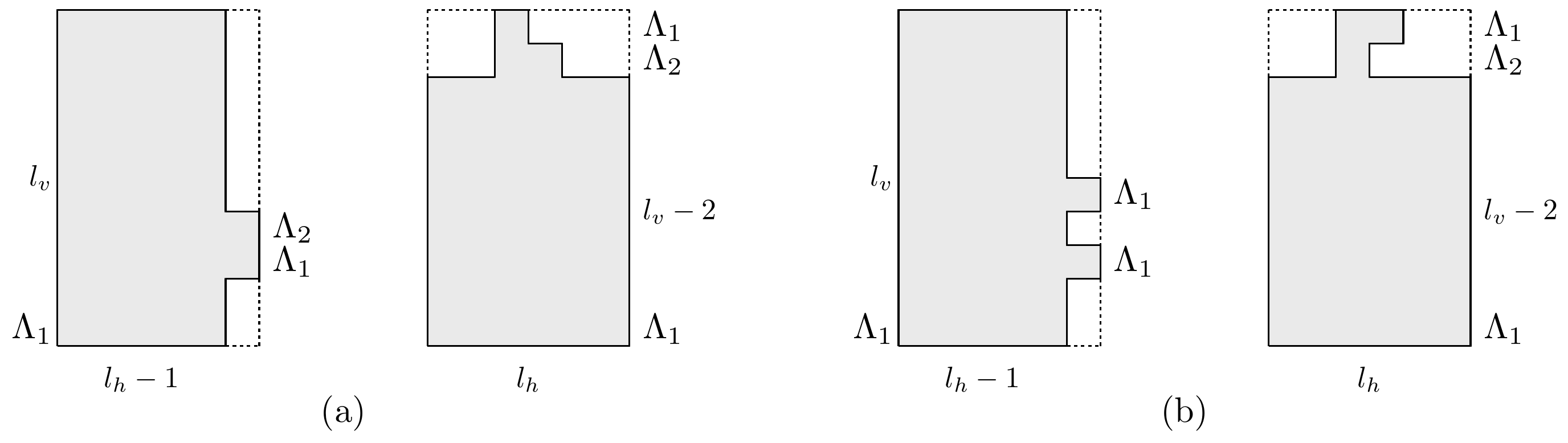} 
    \caption{(a) The shapes of the critical clusters of pluses in parameter region (I). (b) The shapes of the critical clusters of pluses in parameter region (II).}
    \label{fig:enter-label}
\end{figure}

Another form of anisotropy is explored in \cite{nardi1996low, nardi1999ising}, where varying magnetic fields across different spatial regions affect not only the shape of the critical droplet but also the dynamic behavior of the model. The authors of \cite{nardi1996low} consider a magnetic field with alternating signs along different rows of the two-dimensional torus and study the metastable behavior of the system under Glauber dynamics at very low temperature. Specifically, let $\Lambda_1$ (resp. $\Lambda_2$) be the union of the odd (resp. even) rows in $\Lambda=\Lambda_1 \cup \Lambda_2$, they define the Hamiltonian function as follows
\begin{align}
    H(\sigma):=-\frac{J}{2}\sum_{\substack{i,j \in \Lambda \\ d(i,j)=1}} \sigma (i) \sigma (j) 
    -\frac{h_1}{2}\sum_{i \in \Lambda_1} \sigma (i)+\frac{h_2}{2}\sum_{i \in \Lambda_2} \sigma (i),
\end{align}
where $J>0$ and $h_1 \geq h_2 > 0$. 
According to the relations between the three parameters, it is possible to represent the phase diagram as in Figure \ref{fig:phase_diagram}. 
In particular, 
\begin{itemize}
    \item if $h_1>h_2$ and $0<h_2<2J$, then the homogeneous state $\textbf{+1}$ is the stable state.
    \item if $h_1<h_2$ and $0<h_1<2J$, then the homogeneous state $\textbf{-1}$ is the stable state.
    \item if $h_1,h_2>2J$, then the configuration $\sigma_{\nu}$, i.e. the configuration that contains all plus spins on $\Lambda_1$ and all minus spins on $\Lambda_2$, is the stable state.
\end{itemize}
We focus on the two most interesting regions: (I) $2J>h_1>h_2>0$ and $h_2<2J-h_1$, (II) $2J>h_1>h_2>0$ and $h_2>2J-h_1$. See Figure \ref{fig:phase_diagram}. In both these regions, the shape of the critical configurations is non-Wulff, similar to what was observed in \cite{nardi1999ising}. Indeed, the vertical side $l_v$ of the critical rectangle  is significantly longer (almost the double) than the horizontal side $l_h$. By contrast, in the Wulff rectangle, the horizontal side is longer than the vertical side and grows infinitely larger. More precisely, the critical lengths are equal to
\begin{align}
    &l_h := \left \lfloor \frac{2J-h_2}{h_1-h_2} \right \rfloor \qquad \text{ and } \qquad
    l_v:=2l_h-1,
\end{align}
and the critical droplets of the first parameter region have the shapes as in Figure \ref{fig:enter-label}-(a), while the saddles of the second region are configurations with a critical cluster as in Figure \ref{fig:enter-label}-(b).
Specifically, the cluster of pluses in the critical configurations differs from a rectangular shape and exhibits protuberances of specific dimensions, which depend on the region of the parameter diagram being considered. These protuberances play a crucial role in the evolution of the systems, since they facilitate the transition between metastable and stable states, triggering either the growth or the contraction of the cluster by determining the return to the metastable state or the progression toward the stable state.

\section{Long-range Ising model}
\label{subsec:longrange}
The long-range Ising model generalizes the traditional Ising model by allowing spins to influence each other over longer distances. In some cases the nature of these interactions can determine the emergence of new phases of matter or critical points that are not observed in the short-range model. In addition, the presence of long-range interactions makes the study non-local, meaning that one spin can have a significant effect on another spin far away in the lattice, and this non-locality complicates both analytical computations and numerical simulations.

In \cite{van2019nucleation}, the authors consider the long-range Ising model in one dimension. 
In fact, unlike the short-range Ising model, this model undergoes a phase transition already in one dimension, and this phase persists in fairly fast decaying fields, see \cite{cassandro2005geometry, dyson1972existence,bissacot2018contour, littin2017quasi}. 
In low dimensions, such long-range models are known to behave like higher-dimensional short-range models.

In particular, the authors of \cite{van2019nucleation} consider a box $\Lambda \subset \mathbb{Z}$ and they choose free boundary conditions. The energy of the system is defined as follows
\begin{equation}\label{hamiltonian_function_LR}
H(\sigma):=-\sum_{\substack{i,j \in \Lambda}} \mathcal{J}(d(i,j))  \sigma(i) \sigma(j)-h \sum_{i \in \Lambda} \sigma(i),
\end{equation}
where $\mathcal{J}: \mathbb{N} \to \mathbb{R}$ is a positive and decreasing function. The pair interaction $\mathcal{J}$ represents the straight of the interactions between two sites that decreases with the inverse of the distance between them and it can represent two different type of decay: an exponential decay as $\mathcal{J}(n)= J \alpha^{-n+1}$ with $J>0,\alpha>1$, and a polynomial decay as $\mathcal{J}(n)= J n^{-\lambda}$ with $J>0,\lambda>0$.
According to the choice of the decay, the behavior of this model can be equal or different to that of the classical short-range Ising model. Indeed, if the power of the interaction is exponential, then a single plus spin in $\Lambda$ will trigger the nucleation of the stable phase. In the other case, the long-range effects are more visible and the model shows a different evolution.

\begin{figure}[htb!]
\centering
    \includegraphics[scale=0.5]{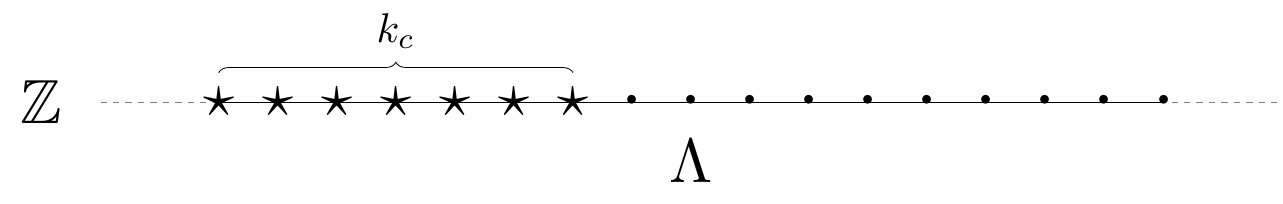}
    \caption{An example of the critical droplet attached to the left side of $\Lambda$. The stars represent the plus spins, the circles are the minus spins.
    }
  \label{1d_Ising}
    \end{figure}

The authors of \cite{van2019nucleation} consider the evolution of the model with Glauber dynamics and they prove that, if the external magnetic field is small enough, than the homogeneous state $\textbf{-1}$ is the unique metastable state and they estimate the transition time from it to the stable state $\textbf{+1}$ (Theorems \ref{thm:asymptotic_result_Metropolis_1}, \ref{thm:asymptotic_result_Metropolis_3}). In order to estimate the mean exit time, they find the size of the critical configuration and they show that it is macroscopic or mesoscopic, according to the value of the external magnetic field. 
In particular, they prove that the nucleation occurs from the boundary due to the choice of the free conditions and the critical droplets are those configurations with $k_c\in \mathbb{N}$ consecutive plus spins on the left (resp. on the right) side and minus spins in the rest of $\Lambda$, see Figure \ref{1d_Ising}. The size $k_c$ depends on the choice of the pair interactions, in particular for $\Lambda$ large enough, we have
\begin{align}
    k_c \simeq
    \begin{cases}
    \left ( \frac{J}{h(\lambda -1)} \right )^{\frac{1}{\lambda-1}} &\text { if } \mathcal{J}(n)= J n^{-\lambda}, \notag \\
        \lceil \log_{\alpha} \left ( \frac{J}{h(1-\alpha^{-1})} \right ) \rceil & \text { if } \mathcal{J}(n)=J \alpha^{-n+1} .
    \end{cases}
\end{align}
We observe that in the second case, with exponential decay of the interaction, the system behaves essentially as the nearest-neighbours one-dimensional Ising model. Indeed, if the distance $n=1$ then $\lim_{\alpha \to \infty} \mathcal{J}(n)=J$, and it is equal to zero in the other cases. Moreover, $k_c=1$ for $h<J$ and $\alpha$ large enough.

In higher dimensions, at the moment, no rigorous results are currently available regarding the metastable behavior of the long-range Ising model. However, an important step for identifying the typical configurations that trigger the nucleation process can be found in \cite{jacquier2024isoperimetric}. Specifically, the authors of \cite{jacquier2024isoperimetric} analyze the two-dimensional long-range Ising model with biaxial interactions (i.e. non-zero interaction along both the horizontal and vertical directions) described by the following Hamiltonian function
\begin{equation}
H^\lambda(\sigma):=-\sum_{x,y \in \Lambda,\atop{x \neq y}} \mathcal{J}(d^\lambda (x,y)) \sigma(x) \sigma(y)-h \sum_{x \in \Lambda} \sigma(x),
\end{equation}
where $\mathcal{J} (d^\lambda(x,y))$ is given by
the fractional bi-axial function,
\begin{equation}
J(d^\lambda(x,y)):= 
\begin{cases}
\frac{1}{d^{\lambda}(x,y)} \qquad & \text{ if either } x_2=y_2 \text{ or } x_1=y_1, \notag \\
0 \qquad & \text{ otherwise,}
\end{cases}
\end{equation}
where $x=(x_1,x_2), \, y=(y_1,y_2)\in \Lambda$, $\lambda>1$, and
\begin{equation}
\frac{1}{d^{\lambda}(x,y)} := \frac{1}{|x_2-y_2|^\lambda}\textbf{1}_{\{ x_1=y_1, \, x_2 \neq y_2\}}  + \frac{1}{|x_1-y_1|^{\lambda}} \textbf{1}_{\{  x_2=y_2, \, x_1 \neq y_1\}}.
\end{equation}
Considering the \emph{nonlocal perimeter} $Per_{\lambda}$, defined as
\begin{align}
    Per_{\lambda}(\mathcal{P}):=\sum_{x \in \mathbb{Z}^2 \cap \mathcal{P}, \, y \in \mathbb{Z}^2 \cap \mathcal{P}^c} \frac{1}{d^{\lambda}(x,y)},
\end{align}
it is possible to rewrite the Hamiltonian function as in \eqref{eq:peierls_hamiltonian}. In this way, the solution of the isoperimetric inequality present in \cite{jacquier2024isoperimetric} enables the identification of minimal-energy configurations for a given fixed magnetization. Specifically, for different value of the magnetization $n$, they find different shape of the critical droplets. In fact, for $n=l^2$ (resp. for $n=l(l+1)$) the configurations with minimal energy contain a cluster of pluses with square (resp. quasi-square) shape. While if $n=l^2+k$ (resp. $n=l(l+1)+k$), then the critical droplets have the same minimal perimeter $4l+2$ (resp. $4l+4$) and one of the following shapes: square (resp. quasi-square) with a protuberance, rectangle, rectangle with a protuberance attached along one of the shorter sides.

\section{Some extensions of the Ising model}\label{sec:extension}
The Ising model has inspired numerous extensions that generalize its framework to describe a wider variety of physical phenomena. In this Section, we will discuss the most prominent ones.

\subsection{Ising model with Kawasaki dynamics}\label{sec:Kawasaki}
\begin{figure}[htb]
        \begin{center}
        \includegraphics[scale=0.4]{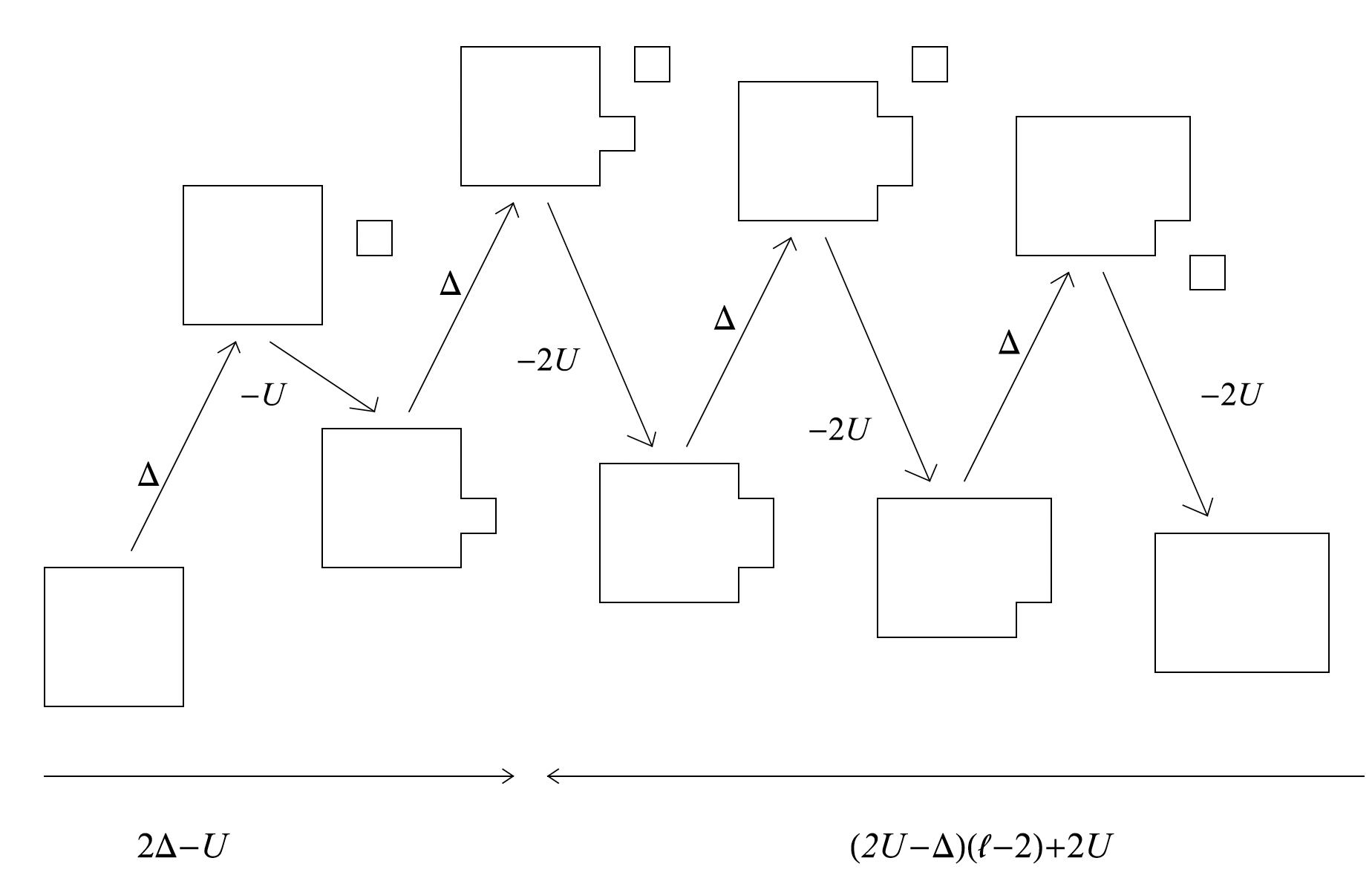}
        \end{center}
        \caption{The growth of a cluster and the energy cost of each step. Specifically, when a particle get in the box the energy increases by $\Delta$, while when the particle attaches to the cluster the energy decreases by $U$ or $2U$ according to the number of the new bonds.}
        \label{fig:Kawasaki_evolution}
\end{figure}

\begin{figure}[!htb]
        \begin{center}
        \includegraphics[scale=0.35]{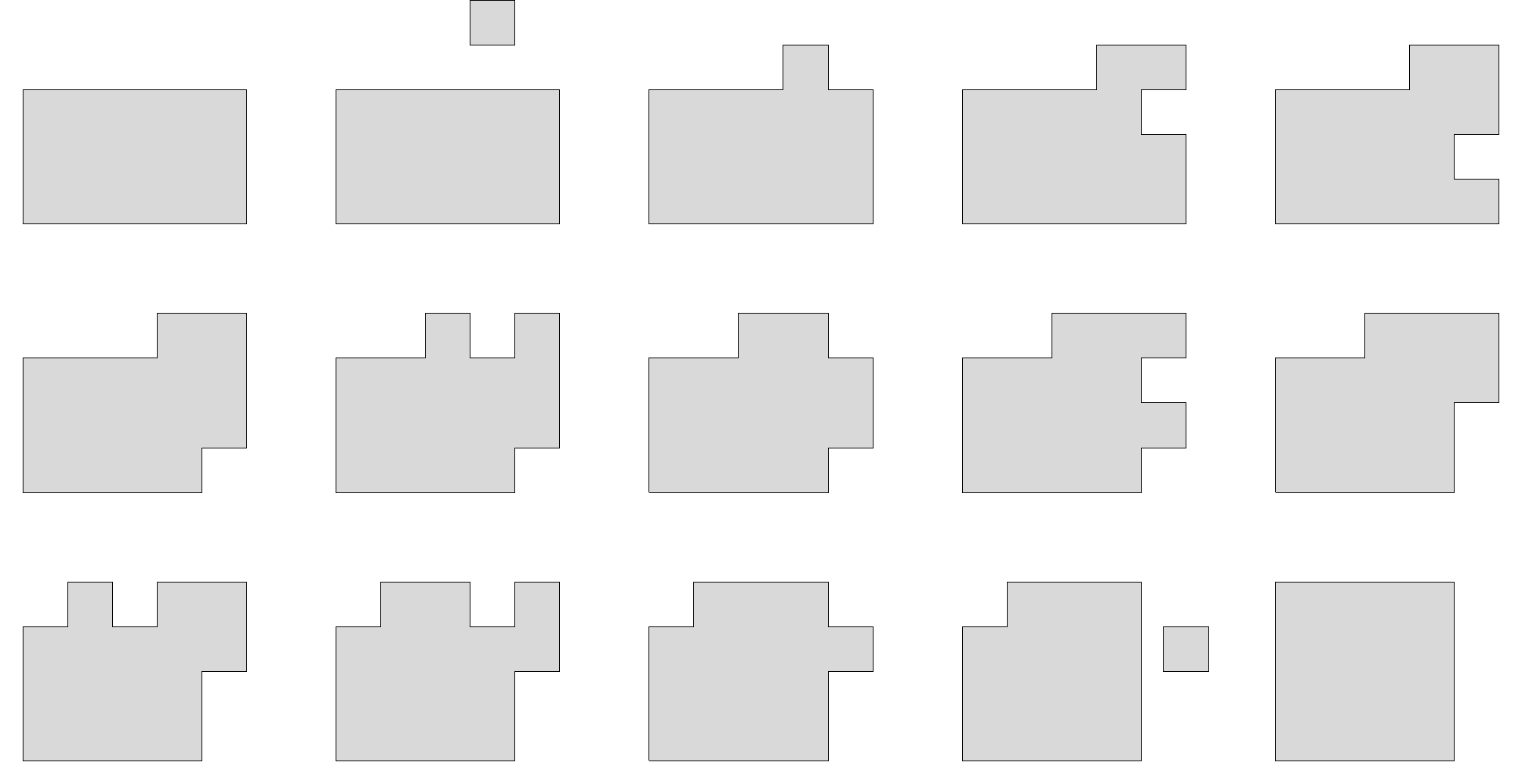}
        \end{center}
        \caption{The movement of the particles along the boundary of the cluster to reshape the rectangular droplet into a square droplet.}
        \label{fig:trenini}
    \end{figure}

We analyze the evolution of the Ising model under a conservative dynamics, Kawasaki dynamics. In contrast to Glauber dynamics, Kawasaki dynamics conserves total magnetization by allowing spin exchanges between neighboring sites. This conservation adds an extra layer of complexity in the formation of the droplets and in their growth. This leads to longer transition times compared to Glauber dynamics. See Figures \ref{fig:Kawasaki_evolution} and \ref{fig:trenini} for two simplified illustrations of the evolution of the system.

In the following, we formally describe the Kawasaki dynamics and reformulate the Hamiltonian function.
Let $\Lambda$ be a large set of vertices and let 
$\partial^- \Lambda:=\{x \in \Lambda \, | \, d(x,y)=1, \,\, \exists  \, y \not \in \Lambda \}$
 and 
$\partial^+ \Lambda:=\{x \not \in \Lambda \, | \, d(x,y)=1, \,\, \exists  \, y \in \Lambda \}$
be the internal and the external boundary of $\Lambda$, respectively. Set $\Lambda^-:=\Lambda \setminus \partial^- \Lambda$, $\Lambda^+:=\Lambda \cup \partial^+ \Lambda$ and let $\Lambda^{*,-}$ be the set of non-oriented bonds in $\Lambda^-$. 
We associate to each site $i \in \Lambda$ a random variable $\sigma(i)$, that takes value in $\{0,1\}$. These values indicate the absence or presence of a particle at the site $i$. Thus, in this case, the configuration space is $\mathcal{X}=\{0,1\}^\Lambda$ and the energy of a configuration $\sigma$ is defined as
\begin{align}\label{H_K}
    H_K(\sigma)=-U \sum_{(i,j) \in \Lambda^{*,-}}\sigma(i)\sigma(j)+\Delta \sum_{i \in \Lambda} \sigma(i)
\end{align}
where $U>0$ is an interaction term corresponding to a \emph{binding energy} for each nearest-neighbor pair of particles in $\Lambda^-$, while $\Delta>0$ represents the \emph{activation energy} for each particle in $\Lambda$. From a  physical point of view, $\Delta \in (0,U)$ corresponds to the unstable gas, $\Delta=U$ to the spinodal point, $\Delta \in (U,2U)$ to the metastable gas, $\Delta=2U$ to the condensation point, and $\Delta \in (2U,\infty)$ to the stable gas. 

We observe that the Hamiltonian function \eqref{H_K} is analogous to the classical Hamiltonian of the Ising model defined in \eqref{hamiltonianFunction}. Indeed, after we make the substitution $\sigma(i)=(\xi(i)+1)/2$ with $\xi(i) \in \{-1,+1\}$, we obtain
\begin{align}
    H_K(\sigma)&=-U \sum_{(i,j) \in \Lambda^{*,-}}\frac{\xi(i)+1}{2}\frac{\xi(j)+1}{2}+\Delta \sum_{i \in \Lambda} \frac{\xi(i)+1}{2} \notag \\
    &=-\frac{U}{4} \sum_{(i,j) \in \Lambda^{*,-}}\xi(i)\xi(j)-\frac{2U-\Delta}{2} \sum_{i \in \Lambda} \xi(i)+ C_\Lambda \notag \\
    &=H(\xi)+ C_\Lambda,
\end{align}
where $C_\Lambda$ is a constant that depends on the size of the box $\Lambda$, and we used $J=U/4$ and $h=(2U-\Delta)/2$.

The Kawasaki dynamics mimics the effect of an infinite gas reservoir outside $\Lambda$ with density $\rho_{\beta}=e^{-\beta \Delta}$
according to the activation energy $\Delta$ in \eqref{H_K}. This dynamics is a standard Metropolis dynamics with an open boundary: along each bond connecting $\partial^- \Lambda$ from the outside, the particles are created with rate $\rho_{\beta}$ and they are annihilated with rate 1, while inside $\Lambda^-$ the particles are conserved and jump at a rate that depends on the change in energy associated with the jump.

Formally,
let $b=(x \rightarrow y)$ be an oriented bond, i.e., an ordered pair of nearest-neighbor sites, we define the sets
\begin{align}
    &\Lambda^{*,orie}=\{ (x \rightarrow y) | \, x,y \in \Lambda\}, \\
    &\partial\Lambda^{*,in}=\{ (x \rightarrow y) | \, x \in \partial^+ \Lambda,y \in \partial^-\Lambda\}, \\
    &\partial\Lambda^{*,out}=\{ (x \rightarrow y) | \, x \in \partial^- \Lambda,y \in \partial^+\Lambda\}.
\end{align}
Two configurations $\sigma,\sigma'\in \mathcal{X}$ with $\sigma \neq \sigma'$ are communicating configurations $\sigma \sim \sigma'$, if there exists a bond $b\in \Lambda^{*,orie} \cup \partial\Lambda^{*,in} \cup \partial\Lambda^{*,out}$ such that $\sigma'=T_b (\sigma)$, where $T_b (\sigma)$ is the configuration obtained from $\sigma$ as follows:
\begin{itemize}
    \item[(i)] $b=(x \rightarrow y) \in \Lambda^{*,orie}$:
    \begin{align}
        (T_b(\sigma))(z)= 
        \begin{cases}
            \sigma(z) & \text{ if } z \neq x,y, \\
            \sigma(x) & \text{ if } z =y, \\
            \sigma(y) & \text{ if } z =x. 
        \end{cases}
    \end{align}
    \item[(ii)] $b=(x \rightarrow y) \in \partial\Lambda^{*,in}$:
    \begin{align}
        (T_b(\sigma))(z)= 
        \begin{cases}
            \sigma(z) & \text{ if } z \neq y, \\
            1 & \text{ if } z =y. 
        \end{cases}
    \end{align}
    \item[(iii)] $b=(x \rightarrow y) \in \partial\Lambda^{*,out}$:
    \begin{align}
        (T_b(\sigma))(z)= 
        \begin{cases}
            \sigma(z) & \text{ if } z \neq x, \\
            0 & \text{ if } z =x.
        \end{cases}
    \end{align}
\end{itemize}

These transitions between $\sigma$ and $T_b(\sigma)$ correspond to particle motion in $\Lambda$, and creation/annihilation of particles in $\partial^-\Lambda$. 
Thus, Kawasaki dynamics is defined to be the continuous-time Markov chain $(\sigma_t)_{t\geq 0}$ on $\mathcal{X}$ with transition rates
\begin{align}
c_{\beta}(\sigma,\sigma')=\textbf{1}_{\sigma \sim \sigma'} \exp{\{-\beta [H_K(\sigma')-H_K(\sigma)]_+\}}
\end{align}
for every $\sigma \neq \sigma'$.
We observe that this dynamics is reversible with respect to the Gibbs measure \eqref{def:gibbs} with \eqref{H_K}.

The first studies of the metastable behavior under Kawasaki dynamics are in 
\cite{hollander2000metastability, den2000nucleation}. There, the authors investigate the metastability and the nucleation for a local version of the 2D lattice gas with Kawasaki dynamics at low temperature and low density. Moreover, they highlight the main differences in the evolution of the system under this conservative dynamics and the Glauber dynamics. One of the main differences is that, under Kawasaki dynamics, rectangular droplets tend to become square through a movement of particles along the border of the droplet, see Figure \ref{fig:trenini}, while
under Glauber dynamics subcritical rectangular droplets tend to shrink along the shortest sides and the supercritical rectangular droplets tend to grow uniformly in all directions.

The 3D version of the same model is  considered in \cite{den2003droplet}. In particular, the authors identify the size and shape of the critical droplet in 3D case and the time of its nucleation in the limit of low temperature and low density.
Additionally, the results present in \cite{bovier2006sharp} offer a thorough examination of the metastability phenomenon, refining the outcomes achieved in \cite{hollander2000metastability} for two dimensions and in \cite{den2003droplet} for three dimensions.

Another intriguing analysis of the metastable behavior involves examining an anisotropic model governed by Kawasaki dynamics, where particles preferentially move along one of the two directions. See \cite{nardi2005anisotropy, baldassarri2022criticalweak, baldassarri2022critical, baldassarri2021metastability} for detail about the evolution of the system with weak and strong anisotropy.

All of these results, for both isotropic and anisotropic case, are comparable to those obtained for the (isotropic and anisotropic) Ising model subject to Glauber dynamics. 
For instance, the findings reported in \cite{baldassarri2023metastability} for the Kawasaki dynamics on the hexagonal lattice align with those derived for the Ising model in \cite{apollonio2021metastability}.

The generalization of this model by introducing a second type of particle is explored in a series of three papers  
\cite{den2012kawasaki, den2012metastability, den2011kawasaki}. These works investigate the two-dimensional lattice gas consisting of two types of particles governed by Kawasaki dynamics at low temperature in a large finite box with an open boundary. In particular, they consider a positive activation energy for each particle that depends on its type, a negative binding energy for each pair of different particles occupying neighboring sites, and no binding energy between particles of the same type. 
Different results are obtained in \cite{jacquier2024particle}, where the authors consider alternative binding energy and zero-boundary conditions. 

A more recent series of papers \cite{gaudilliere2009ideal, baldassarri2024droplet, baldassarri2024homogeneous} deals with the evolution of a lattice gas under Kawasaki dynamics at inverse temperature $\beta>0$ in a large finite box $\Lambda_\beta \subset \mathbb{Z}^2$ whose volume depends on $\beta$.
There the authors show that the subcritical droplets behave as quasi-random walks, and they analyze how subcritical droplets form and dissolve on multiple space–time scales when the volume is either moderately large, i.e. $|\Lambda_\beta|=e^{\Theta_1\beta}$ with $\Theta_1 \in (\Delta,2\Delta-U)$, or very large, i.e. $|\Lambda_\beta|=e^{\Theta_2\beta}$ with $\Theta_2 \in (\Delta,\Gamma-(2\Delta-U))$ where $\Gamma$ is the energy of the critical droplets in the local model.

\subsection{Blume Capel model}\label{blume_capel}

The Blume-Capel model is an extension of the Ising model that incorporates a third possible state for the spins on a lattice, i.e. $\mathcal{X}=\{-1,+1,0\}^\Lambda$, offering a richer and more versatile framework for studying metastability. It was originally introduced to study the $\prescript{3}{}{He}-\prescript{4}{}{He}$ phase transition and 
it may be interpreted a system of particles with spin. At a given site $i$, the spin value 
$\sigma(i)=0$ represents the absence of a particle (a vacancy), while $\sigma(i)=\pm 1$ indicates the presence of a particle with either positive or negative spin at that site on the lattice.
The Hamiltonian for the Blume-Capel model is given by
\begin{align}
H(\sigma)
&:=J \sum_{\substack{i,j \in \Lambda \\ d(i,j)=1}} (\sigma(i)-\sigma(j))^2-\lambda \sum_{i\in \Lambda} \sigma(i)^2-h\sum_{i\in \Lambda} \sigma(i),
\end{align}
where $J>0$ is the coupling constant, $h$ is the external magnetic field, and $\lambda$ is an anisotropy parameter, called \emph{chemical potential}, which assigns an energy penalty or preference for the zero-state. This term introduces competition between ordered ($\pm$ \textbf{1}) and disordered ($\textbf{0}$) phases.

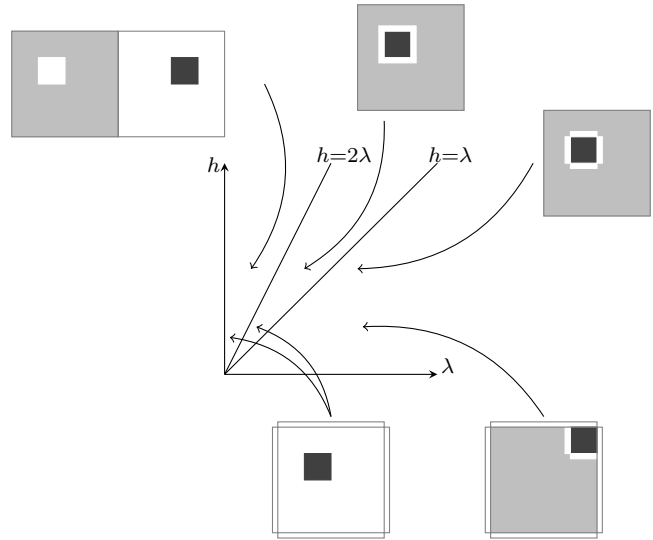
\begin{figure}[hbt!]
\begin{center}
\begin{tikzpicture}[scale=0.7]
  \draw[-stealth](4,4)--(8,4);
  \draw[-stealth](4,4)--(4,8);
  \draw (4,4)--(8,8);
  \draw (4,4)--(6,8);
  \node[] at (6.,8.2) {${\phantom{m}_{h=2\lambda}}$};
  \node[] at (8.,8.2) {${\phantom{m}_{h=\lambda}}$};
  \node[] at (3.55,8.) {${\phantom{m}_{h}}$};
  \node[] at (7.95,4.2) {${\phantom{m}_{\lambda}}$};
  [very thick]
  \draw[->] (4.75,9.5) to [bend left] (4.5,6.);
  \draw[->] (7.,8.8) to [bend left] (5.5,6.);
  \draw[->] (9.8,8.) to [bend left] (6.5,6.);
  \draw[->] (6.,3.2) to [bend right] (4.1,4.7);
  \draw[->] (6.,3.2) to [bend right] (4.6,4.9);
  \draw[->] (10.,3.2) to [bend right] (6.6,4.9);

  \draw[fill=lightgray,lightgray] (0,8.5) rectangle (2,10.5);
  \draw[gray] (0.,8.5) rectangle (2.,10.5);
  \draw[fill=white,white] (0.5,9.5) rectangle (1.,10);
  \draw[fill=white,white] (2.5,8.5) rectangle (4.5,10.5);

  \draw[gray] (2.,10.5) rectangle (4.,8.5);
  \draw[fill=darkgray,darkgray] (3.0,9.5) rectangle (3.5,10);

  \draw[fill=lightgray,lightgray] (6.5,9.) rectangle (8.5,11.);
  \draw[gray] (6.5,9.) rectangle (8.5,11.);
  \draw[fill=darkgray,darkgray] (7.0,10.) rectangle (7.5,10.5);
  \draw[fill=white,white] (6.9,10.5) rectangle (7.6,10.6);
  \draw[fill=white,white] (6.9,9.9) rectangle (7.,10.5);
  \draw[fill=white,white] (7,9.9) rectangle (7.6,10.);
  \draw[fill=white,white] (7.5,10) rectangle (7.6,10.5);

  \draw[fill=lightgray,lightgray] (10.,7.) rectangle (12.,9.);
  \draw[gray] (10.,7.) rectangle (12.,9.);
  \draw[fill=darkgray,darkgray] (10.5,8.) rectangle (11.,8.5);
  \draw[fill=white,white] (10.5,8.5) rectangle (11.,8.6);
  \draw[fill=white,white] (10.5,7.9) rectangle (11.,8.);
  \draw[fill=white,white] (10.4,8.) rectangle (10.5,8.5);
  \draw[fill=white,white] (11.,8.) rectangle (11.1,8.5);

  \draw[fill=white,white] (5,1.) rectangle (7,3.);
  \draw[fill=darkgray,darkgray] (5.5,2.) rectangle (6.,2.5);
  \draw[gray] (5.,3.) rectangle (7.,3.1);
  \draw[gray] (5.,0.9) rectangle (7.,1.);
  \draw[gray] (4.9,1.) rectangle (5.,3.);
  \draw[gray] (7.,1.) rectangle (7.1,3.);

  \draw[fill=lightgray,lightgray] (9.,1.) rectangle (11.,3.);
  \draw[fill=darkgray,darkgray] (10.5,2.5) rectangle (11.,3.);
  \draw[fill=white,white] (10.5,2.4) rectangle (11.,2.5);
  \draw[fill=white,white] (10.4,2.5) rectangle (10.5,3.);
  \draw[gray] (9.,3.) rectangle (11.,3.1);
  \draw[gray] (9.,0.9) rectangle (11.,1.);
  \draw[gray] (8.9,1.) rectangle (9.,3.);
  \draw[gray] (11.,1.) rectangle (11.1,3.);
\end{tikzpicture}
\end{center}
    \caption{Schematic representation present in \cite{cirillo2024homogeneous} of the metastable behavior of the Blume-Capel model with periodic boundary condition (top pictures) and with zero-boundary condition (bottom pictures). The minus spins are represented by gray color, plus spins are in black and zero spins in white.}
  \label{BC_critical}
    \end{figure}

The metastable behavior of the Blume-Capel model with Glauber dynamics has been studied for the first time in \cite{cirillo1996metastability}. In this work, the authors investigate the first excursion from the metastable state $\textbf{-1}$ and the stable state $\textbf{+1}$ at low temperature. More precisely, they consider a large finite volume with periodic boundary conditions and they compute the asymptotic behavior of the transition time in the parameter region $h>\lambda>0$. Moreover, they describe the typical tube of trajectories during the transition and show that the mechanism of transition changes when the line $h=2\lambda$ is crossed.
In particular, for $h>2\lambda>0$, the system during the transition from $\textbf{-1}$ to $\textbf{+1}$ crosses the homogeneous state $\textbf{0}$. Indeed, a suitable critical droplet of zeros starts to grow in a sea of minuses until it covers the whole volume, then the nucleation of a critical droplet of pluses takes place in a sea of zeros until the system reaches the stable state. 
While for $2\lambda>h>\lambda>0$, the plus phase is created directly in the minus phase via the formation of a suitable critical nucleus, i.e. a plus square droplet separated from the sea of minuses by a layer of zeros of width one. See the top pictures in Figure \ref{BC_critical}.
Similar results are obtained in \cite{manzo2001dynamical} for the infinite volume case. 

The metastability for the Blume-Capel model with zero chemical potential is analyzed in \cite{landim2016metastability, landim2019metastability}, where the authors characterize the set of the critical configurations and obtain sharp estimates for the transition time at low temperature regime.
In \cite{cirillo2017sum}, the authors derive, with a different method, the same result proven in \cite{landim2016metastability} on the sharp estimate of the exit time from the metastable state in the zero chemical potential case. 

For a specific choice of parameters, the Blume-Capel model exhibits multiple metastable states that are not degenerate in energy. This case is analyzed in \cite{cirillo2013relaxation}, where sufficient conditions are provided to identify multiple metastable states, and leveraging these results, the transition time is estimated in probability and the set of critical configurations is identified.

The evolution of the Blume-Capel model with zero-boundary condition is explored under Glauber dynamics in a recent work \cite{cirillo2024homogeneous}, and under Kawasaki dynamics in \cite{jacquier2024particle}. See the bottom pictures in Figure \ref{BC_critical} for the shapes of the critical configurations of Blume-Capel model with zero-boundary under Glauber dynamics.
Another study on the metastability for this model is in \cite{kim2021metastability}, where the author assume $h=\lambda=0$ and he highlights that no critical saddle configurations exist and the transition starting from a homogeneous configuration must occur along a massive flat plateau of saddle configurations to reach another monochromatic state.

\subsection{Potts model}\label{sec:Potts}
Another generalization of the Ising model is the Potts model where the values of the spins are more than two. Specifically, the spins of this model may take $q$ different values and in the $q$-state Potts model each spin lies on a vertex of a finite two-dimensional rectangular lattice $\Lambda=(V,E)$ where $V=\{0,...,K-1\} \times \{0,...,L-1\}$ is the set of vertices and $E$ is the set of pairs of vertices at distance one from each other. To each configuration $\sigma \in \{1,...,q\}^{V}$ it is associated the following energy function:
\begin{align}
    H(\sigma):=&-J \sum_{\substack{i,j \in E}} \textbf{1}_{\{\sigma(i)=\sigma(j)\}}-h\sum_{i\in V} \sigma(i).
\end{align}
The Ising model can be seen as a special case of the Potts model with $q=2$. While for $q>2$, the Potts model introduces greater complexity, allowing the study of systems with more possible states per site. This makes it a powerful tool for analyzing phase transitions in more intricate systems than those described by the Ising model.
 
The initial studies on the order of the phase transitions and metastability for Potts model are present in \cite{peruggi1983potts} by considering Bethe lattices, and in \cite{de1991metastability} on the Cayley Tree.

Two more recent works on the analysis of the metastable behavior of the Potts model are \cite{bet2022metastability, bet2023metastability}. In both, the authors assume periodic boundary conditions and analyze the evolution of the system at low temperature under Glauber dynamics. \cite{bet2022metastability} focuses on the case $h<0$, which is characterized by the presence of $q-1$ stable configurations and only unique metastable state. On the contrary for the case $h>0$, studied in \cite{bet2024metastability}, there exists only one stable state and $q-1$ metastable configurations. A complete description of the metastability is provided in both cases. In fact, the authors study the asymptotic behavior of the first hitting time from the set of the metastable states to the set of the stable states (Theorems \ref{thm:recurrence_property_Metropolis}, \ref{thm:asymptotic_result_Metropolis_1}, \ref{thm:asymptotic_result_Metropolis_3}), determine the exponent of the mixing time and an estimate for the spectral gap (Theorem \ref{thm:mixing_time_spectral_gap_Metropolis}), and describe the critical configurations and the tube of typical trajectories in both cases. Moreover, in \cite{bet2022metastability}, the prefactor of the expected value of the transition time is estimated (Theorem \ref{thm:mean_crossover_time}).

There are many works about the analysis of the metastability for Potts model without external magnetic field \cite{bet2021critical, nardi2019tunneling, kim2022metastability, kim2024energy}.
In this setting, the metastable states are not interesting since they do not have a clear physical interpretation, hence the authors focus the attention on the tunneling behavior between stable configurations.
In \cite{nardi2019tunneling} the authors study the transition from any stable configuration to any/some other stable configuration and they derive the asymptotic behavior of the first hitting time and obtain an estimate for the spectral gap. This work is concluded in \cite{bet2021critical}, where the investigation on the transition is extended to the case in which the system, starting from any stable configuration, reaches other stable configuration under the constraint that the path followed does not intersect other stable configurations. Furthermore, they describe the set of minimal gates and the tube of typical paths. 

\cite{kim2022metastability} and \cite{bet2021critical} are contemporaneous works that, despite being developed independently, yield similar results while employing different methodologies. Specifically, \cite{kim2022metastability} investigates the energy landscape and the metastable behavior of Potts models on two-dimensional square or hexagonal lattices at low-temperature with $h=0$. The same results as in \cite{bet2021critical} are achieved, with the main distinction being the replacement of the trajectory tube present in \cite{bet2021critical}  by sharp estimates for the tunneling time (Theorem \ref{thm:mean_crossover_time}) derived in \cite{kim2022metastability} by applying the potential-theoretic approach and extending the results of \cite{nardi2019tunneling}.

Furthermore, the authors of \cite{bet2023metastability} examine a peculiar Potts model similar to the Blume-Capel model (with $\lambda=h=0$) for the number of the spin values and the choice of the coupling constants which establishes a symmetry in the model with respect to two different spins. In fact, let $S=\{1,...,q\}$, they fixed the number of the spin values $q=3$ and they defined the Hamiltonian as follows,
\begin{align}
    H(\sigma):=&-\sum_{x\in S} J_{x,x} \sum_{\substack{i,j \in E}} \textbf{1}_{\{\sigma(i)=\sigma(j)=x\}}+
    \sum_{\substack{x,y\in S \\ x<y} } J_{x,y} \sum_{\substack{i,j \in E}} \textbf{1}_{\{\{\sigma(i),\sigma(j)\}=\{x,y\}\}},
\end{align}
where the coupling constant $J_{x,y}$ are such that $J_{1,1}>J_{2,2}=J_{3,3}$ and $J_{1,2}=J_{1,3}$. With this assumption, 
the spin $1$ is more stable than spins $2, 3$, and the Hamiltonian is symmetric with respect to the spin exchange $2,3$. The results of \cite{bet2023metastability} concern a parameter region where the system exhibits two symmetric metastable states $m_1,m_2$ and a stable one $s$ and the energy barrier between $m_1,m_2$ is equal to that between one of them and $s$. They obtain results on the transition time, on the shapes of the critical configurations, on the mixing and the spectral gap, when the system evolves with Glauber dynamics and the temperature tend to zero.

\subsection*{Funding}
V.J. is funded by the Vidi grant VI.Vidi.213.112 from the Dutch Research Council.

\subsection*{Acknoledgements} 
V.J. is grateful for the partial support of “Gruppo Nazionale per l’Analisi Matematica, la Probabilit\`a e le loro Applicazioni” (GNAMPA-INdAM).

\bibliographystyle{abbrv}
\bibliography{references}

\end{document}